\documentclass[conference]{iscatemplate/IEEEtran}
\usepackage{cite}
\usepackage{amsmath,amssymb,amsfonts}
\usepackage{algorithmic}
\usepackage{graphicx}
\usepackage{textcomp}
\usepackage{xcolor}
\usepackage{fancyhdr}
\usepackage[hyphens]{url}
\usepackage{subcaption}

\usepackage{array}

\def\BibTeX{{\rm B\kern-.05em{\sc i\kern-.025em b}\kern-.08em
    T\kern-.1667em\lower.7ex\hbox{E}\kern-.125emX}}

\fancypagestyle{firstpage}{
  \fancyhf{}

  \fancyfoot[C]{\thepage}
}  

\newcommand{\asplossubmissionnumber}{640}

\usepackage[normalem]{ulem}
\usepackage{tikz}
\usetikzlibrary{calc}
\usepackage{comment}
\usepackage{outlines} %
\usepackage{hyperref} %
\usepackage{booktabs} %
\usepackage{array}    %
\usepackage{amsmath}

\usepackage{layouts}
\usepackage{graphicx}

\usepackage{listings}
\usepackage[linesnumbered,algoruled,boxed,lined]{algorithm2e}
\definecolor{codeblue}{rgb}{0.2,0.2,1}
\definecolor{codegreen}{rgb}{0,0.6,0}
\definecolor{codegray}{rgb}{0.5,0.5,0.5}
\definecolor{codepurple}{rgb}{0.58,0,0.82}
\definecolor{backcolour}{rgb}{0.95,0.95,0.92}
\lstdefinestyle{mystyle}{
    commentstyle=\color{codegreen},
    keywordstyle=\color{codeblue}\bfseries,
    numberstyle=\tiny\color{codegray},
    stringstyle=\color{codepurple},
    basicstyle=\footnotesize\ttfamily,
    breakatwhitespace=false,         
    breaklines=true,                 
    captionpos=b,                    
    keepspaces=true,                 
    numbers=left,                    
    numbersep=5pt,                  
    showspaces=false,                
    showstringspaces=false,
    showtabs=false,
    frame=single,
    tabsize=4,
    columns=flexible
}

\newcolumntype{C}[1]{>{\centering\arraybackslash}p{#1}} %
\newlength{\colw}
\iftrue  %
    \newcommand{\NOTE}[3]{{\color{#1}{[}{~#2:~#3~}{]}}}
\else
    \newcommand{\NOTE}[3]{}
\fi

\begin{document}

\title{Compiling Halide Programs to Push-Memory Accelerators}

\author{
    \IEEEauthorblockN{Qiaoyi Liu, Dillon Huff, Jeff Setter, Maxwell Strange, Kathleen Feng, Kavya Sreedhar, Ziheng Wang,\\ Keyi Zhang, Mark Horowitz, Priyanka Raina, and Fredrik Kjolstad}
    \IEEEauthorblockA{Stanford University
    \\\{joeyliu, dhuff, setter, mstrange, kathleen.feng, skavya, zihengw, keyi, horowitz, praina, kjolstad\}@stanford.edu}
}

\maketitle

\thispagestyle{firstpage}
\pagestyle{plain}

\begin{abstract}
Image processing and machine learning applications benefit tremendously from hardware acceleration, but existing compilers target either FPGAs, which sacrifice power and performance for flexible hardware, or ASICs, which rapidly become obsolete as applications change. Programmable domain-specific accelerators have emerged as a promising middle-ground between these two extremes, but such architectures have traditionally been difficult compiler targets.

The main obstacle is that these accelerators often use a different memory abstraction than CPUs and GPUs: push memories that send a data stream from one computation kernel to other kernels, possibly reordered. To address the compilation challenges caused by push memories, we propose that the representation of memory in the middle and backend of the compiler be altered to combine storage with address generation and control logic in a single structure---a unified buffer. We show that this compiler abstraction can be implemented efficiently on a programmable accelerator, and design a memory mapping algorithm that combines polyhedral analysis and software vectorization techniques to target our accelerator.

Our evaluation shows that the compiler supports programmability while maintaining high performance. It can compile a wide range of image processing and machine learning applications to our accelerator with $4.7\times$ better runtime and $4.3\times$ better energy-efficiency as compared to an FPGA.

\end{abstract}

\section{Introduction}

Image processing and machine learning applications
benefit tremendously from hardware acceleration,
but existing compilers either target FPGAs~\cite{chugh2016dsl, li2020heterohalide}, which
sacrifice power and performance by using flexible hardware,
or directly compile applications to ASICs~\cite{catapult_hls}, which
rapidly become obsolete as applications
change. Programmable domain-specific
accelerators like those shown in~\autoref{tab:accelerator-memory-area-power}
avoid these issues, but have historically been difficult compiler targets.

A key challenge is that these
accelerators use a different memory abstraction than
CPUs and GPUs.  In their execution models, data streams
through the execution units, and the memory units 
push a (possibly reordered)
data stream from one computation kernel to other kernels~\cite{buffets}.
This type of storage is often referred to as a push memory, since it pushes data to the computational units instead of passively waiting for reads.
Since the push memories control both temporary storage and the flow of data, they account for a large fraction of the chip area and power in domain-specific accelerators, as shown in \autoref{tab:accelerator-memory-area-power}. Therefore, these accelerators typically use push memory structures optimized for specific applications, or classes of applications, to minimize area and energy.

These memory optimizations force the compiler to target a different memory abstraction for every application. We address this problem by creating a new push memory abstraction, which we call a \textit{unified buffer} 
since it unifies push buffers. 
It enables efficient push memories by 
bundling control and
address generation with storage, allowing both the compiler and hardware generator
to create more optimized solutions. We also describe a compiler that can harvest these efficiencies by
translating applications that assume a simple von Neumann model of computation into complex data flow streams connected through optimized unified buffers.

Our compiler compiles statically analyzable stencil and dense linear algebra programs expressed in Halide \cite{ragan2013halide} to push memory accelerators.
The design is based on
a simple observation: successful compilers
refine a program from a high-level,
coarse-grained description to a low-level, fine-grained
description. This obvious statement has a profound
implication for compiling to programmable push memory accelerators:
if the target hardware contains more complex
push memory primitives,
then every stage in the compiler that deals with memory must
become more coarse-grained.

In particular, we propose that 
the representation
of memory in the compiler
must be altered to combine storage,
address generation, and control logic in
a single structure---the
unified buffer. Unified buffers serve as an interface inside the compiler between the
application and the architecture. They define both the intermediate representation (IR) used by
the compiler during memory mapping
and the logical behavior that the
hardware architects must implement.
Specifically, our contributions are:
\begin{itemize}
\item A compiler abstraction of push memories, called a unified
buffer, that represents data storage,
address generation, and control in the same
structure.
\item A memory primitive, called a physical unified buffer, that efficiently implements unified buffers on programmable accelerators.
\item A memory mapping algorithm that combines polyhedral analysis with software vectorization to translate unified buffers into configurations of physical unified buffers.
\item An evaluation of our compiler that shows that it can compile a wide range of programs to a programmable accelerator with physical unified buffers and obtain superior  performance and energy-efficiency compared to FPGAs on image processing and machine learning applications.
\end{itemize}

In this paper we give
an overview of our architecture and compiler,
describe the unified buffer abstraction that
serves as an interface between the compiler
and the architecture, explain the phases of our
compiler in detail, and evaluate our compiler
and architecture on several applications
compared to an FPGA.

\begin{table}[tb]
    \caption{
        Programmable domain-specific accelerators typically have push memories, and these account for a significant percentage of chip area and power.
        \label{tab:accelerator-memory-area-power}
    }
    \centering
    \begin{tabular}{p{1.5cm} p{2.5cm} p{0.8cm} p{0.8cm} p{1.0cm}} \toprule
         Domain & Accelerator & Memory Type & Area & Power \\ \midrule
         DNN & TPU \cite{jouppi2017datacenter} & Push & 37\% & N/A \\
         DNN & Eyeriss \cite{chen2016eyeriss} & Push & 67\% & 36--44\% \\
         DNN & Simba PE Array \cite{Shao:2019:SSD:3352460.3358302} & Push & 41\% & 56\% \\
         Sparse DNN & EIE \cite{Han:2016:EEI:3001136.3001163} & Push & 93\% & 59\% \\
         Multiple & Plasticine \cite{plasticine} & Push & 30.2\% & N/A \\
         \bottomrule
    \end{tabular}
\end{table}

\section{Overview}

\begin{figure}
    \centering
    \includegraphics[width=\linewidth,keepaspectratio]{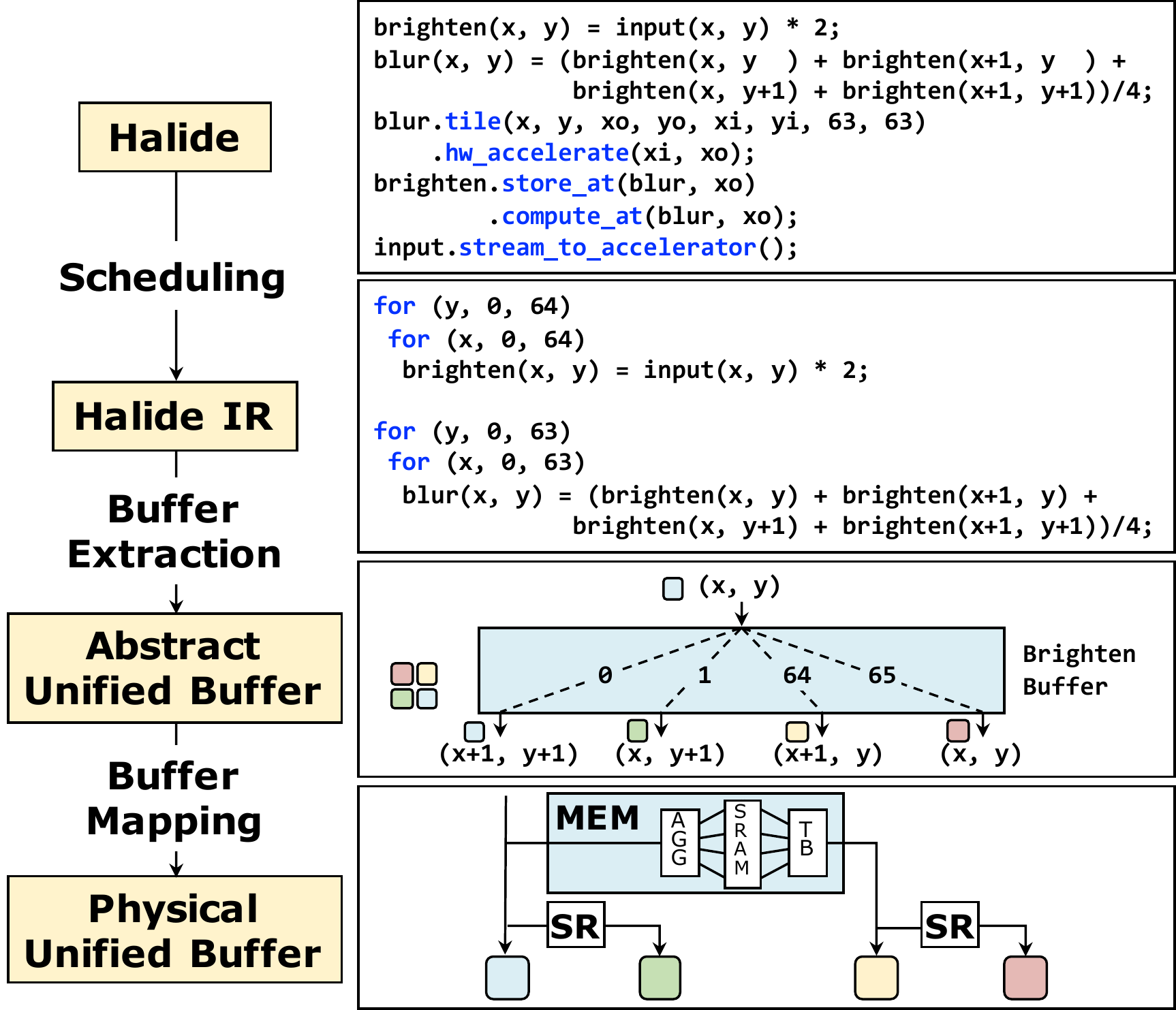}
    \caption{
        Four stages in our system for an example program brighten and blur. The Halide frontend specifies a memory hierarchy (through \texttt{tile}, \texttt{compute\_at} and \texttt{store\_at}) that compiles to loops. The input buffer is eliminated, while the brighten buffer becomes a unified buffer. It is optimized and mapped to shift registers (SR) and our custom vectorized memory tile (MEM).
        \label{fig:brighten_overview}
    }
\end{figure}

We have extended the Halide compiler with a new backend that targets programmable push memories. Our push memory backend takes as input the Halide IR~\cite{ragan2013halide} after schedules have been applied to determine the loop ordering and where intermediate values are stored. \autoref{fig:brighten_overview} shows an overview of the full compiler pipeline from Halide programs to programmed compositions of physical unified buffers. There are three main steps: scheduling, buffer extraction, and buffer mapping. The scheduling step lowers Halide programs to scheduled Halide IR. This step is not new to our work, except that we have extended it with scheduling commands to target a coarse-grained reconfigurable array (CGRA) architecture with programmable push memories.

Our key contribution is the unified buffer abstraction that enables the push memory compiler backend to be broken into two key tasks: extract unified buffers from the application and then map them to concrete hardware implementations. A unified buffer is an abstraction of a buffer that is described only in terms of its input and output ports. Each port is specified not by its implementation, but by a polyhedral specification that describes what data moves through the port and when. The unified buffer cleanly separates the frontend of the compiler, which is concerned with analyzing data movement through the application, from the backend that is hardware-specific and that searches for an implementation that satisfies the port specifications.

The buffer extraction step extracts unified buffers from the Halide IR. This involves polyhedral techniques that determine the necessary ports, summarize the statement instances that use each port and the values they write to or read from the port, and calculate a map from those instances to the times when they use the port.
The buffer mapping step takes as input an abstract unified buffer specification and derives an efficient implementation. The implementation takes in values on the input port at the specified times, and stores them until the time when the specification requires them to be emitted on each output port. Buffer mapping calculates these storage durations from the port specifications, and it may combine registers, SRAMs, and other hardware that we describe in~\autoref{sec:unified-buffer-implementation} to implement the unified buffer.

\section{The Unified Buffer Abstraction}
Since the unified buffer separates the part of the compiler that analyzes the program to determine how values flow through memory from the part that is concerned with creating physical memories to implement that data movement, it has two objectives:
\begin{enumerate}
    \item provide a precise description of the requirements of a push memory at its interfaces, and
    \item maximize opportunities for independent
    optimization on each side of the interface:
    the compiler and the architecture.
\end{enumerate}
The first objective maintains the functionality of the application,
while the second is needed to compile to
efficient implementations. Since push memories
are fundamentally defined by the streams they accept and
generate, we chose to define a unified buffer
by the specification of its I/O streams. 

Exhaustively listing
the values that appear on each port during any real sized
program would require enormous amounts of storage, so
a compact representation of values and the times
when they appear is required. For this we use the
polyhedral model, which provides a well-studied, compact
way to represent schedules and memory access patterns
as integer sets and relations.

\begin{figure}
    \centering
    \includegraphics[width=\linewidth,keepaspectratio]{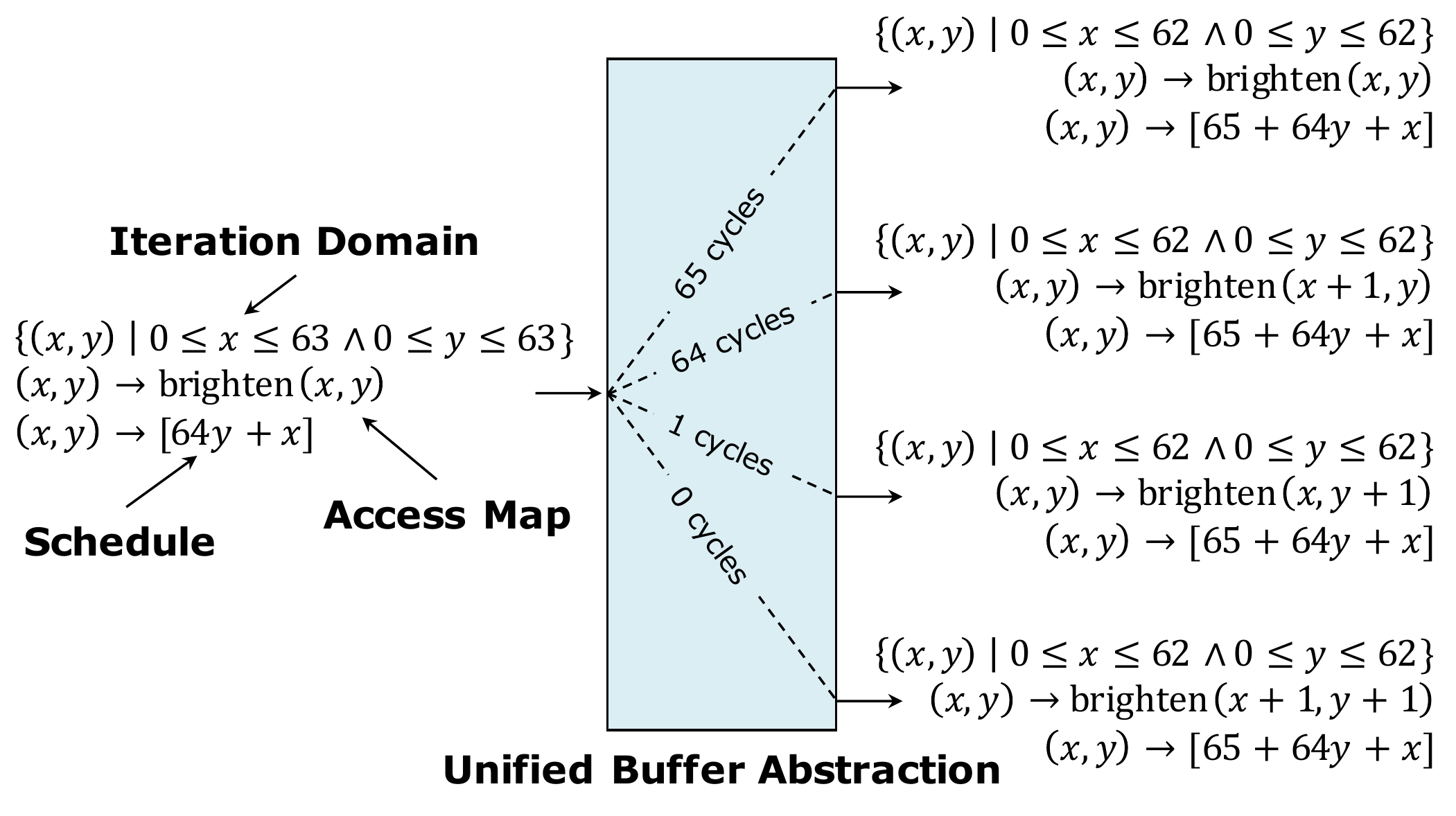}
    \caption{The unified buffer abstraction for the data communication between the brighten and blur functions in the example from \autoref{fig:brighten_overview}. Each port is specified by the polyhedral iteration domain and access map that describe the data written to or read from the buffer, as well as a schedule that describes when those values arrive at the port. From the port specifications, the internal route each data item takes to the ports is calculated, including the distance it travels in cycles.}
    \label{fig:ubuffer_spec}
\end{figure}

\autoref{fig:ubuffer_spec} shows the unified
buffer that is generated to support communication
between the brighten and blur stages of the
example in \autoref{fig:brighten_overview}.
This buffer
accepts one pixel each cycle from the brighten
compute kernel and delivers a $2\times2$ window of pixels
each cycle (after an initial startup delay)
to the blur compute kernel.

To accommodate this bandwidth, the unified
buffer has 5 ports: 1 input port and 4
output ports. The data stream on each port is specified with three
pieces of information:
\begin{itemize}
    \item \textbf{The iteration domain} of the statement instances (operations) that use the port.
    \item \textbf{The access map} for all operations. This function maps points of the iteration domain to the value they read or write on the port.
    \item \textbf{The schedule of all operations} in the iteration domain. This schedule is cycle-accurate as it specifies the exact time in cycles after reset when each operation occurs.
\end{itemize}
These integer sets and relations are
implemented using the polyhedral analysis tool ISL \cite{verdoolaege2010isl}.
For our input port the iteration domain is the set:
\begin{align*}
	\{ (x, y) \mid 0 \leq x \leq 63 \land 0 \leq y \leq 63 \}
\end{align*}
Since the brighten operation, which is the only user of that
port, is surrounded by a two-dimensional loop, the iteration domain
has two index variables, $x$ and $y$. In the above example $y$ is the outermost variable, while $x$ is the innermost variable.

The unified buffer must not just specify what operations use a port. To
synthesize address generation code and optimize memory sizes,
it must also specify what memory locations in the buffer are accessed
by those operations. To specify these memory locations, each port has
an access map. For example the brighten buffer's second output port
has the access map $(x, y) \rightarrow \text{brighten}(x+1, y)$, which means the accessed value is the one to the right of the point in the iteration space of the operation. The other output ports have 
slightly different maps, allowing them to collectively fetch the
required $2\times2$ stencil. 

Conventional polyhedral schedules, such as those produced
by Feautrier's algorithm \cite{feautrier1992some} or PLUTO \cite{bondhugula2008practical}, map
elements of the iteration domain to multidimensional timestamps. Thus, they effectively map the original loops in the program to a new set of loops that implement the new timestamps.
The schedules used by the unified buffer, however, do not map
from one sequence of loop nests to another. Instead, they map
from a sequence of loop nests to cycle times in a hardware design.
That is, they map the operations of the multidimensional iteration domain of the Halide program
to scalar values that represent the \textit{number of cycles after
reset when the operation begins}. Since the hardware design is pipelined, several operations will have the same timestamp. So the schedule for the input port is the integer
function:
\begin{equation}
\label{eq:schedule_example}
(x, y) \rightarrow 64y + x
\end{equation}
This function indicates that the first
brighten operation (and thus the first write to the brighten buffer input port), at coordinate $(0, 0)$, happens $64*0 + 1*0 = 0$ cycles after execution begins, and that
the second brighten operation, at coordinate $(1, 0)$, happens after
$64*0 + 1*1 = 1$ cycle. Furthermore, the output ports emit their first value after $65 + 64*0 + 0 = 65$ cycles, which is the time it takes the first value to travel through the unified buffer. Note that these timestamps impose a requirement that the exact timing of all operations be known, and thus stalls and variable-latency execution are not permitted. Our system respects this limitation by buffering tiles of data required by the accelerator into a large memory (global buffer shown in Figure~\ref{fig:soc}) before running the accelerator. In this way, our system schedules the timestamps of operations within the accelerator, and allows for ready/valid interfaces at boundaries using the global buffer.

Taken together, the unified buffer interface describes
\textit{the observed behavior of the memory at its interface in terms of the operations in the original program}. However, the unified buffer does not
specify the \textit{internal implementation} of
this behavior. Only externally visible scheduling
and binding decisions are expressed. Crucially,
the physical capacity of the memory and the physical
mapping of pieces of data to locations in memory
are omitted. This produces a precise specification
in terms of familiar data structures for a compiler---relations and sets encoded in the polyhedral model---and leaves the architects considerable room to
optimize the design. Next, we discuss how architects can
exploit this interface to design a high-performance,
programmable push memory for dense linear algebra applications.

\section{Physical Unified Buffers}
\label{sec:unified-buffer-implementation}

A physical unified buffer implementation contains the sequencing hardware and storage required to implement a unified buffer abstraction. Since each implementation has a finite capacity and number of ports, buffer implementations support being chained together to create larger capacity memories, higher bandwidth memories, or both. Each implementation may also restrict the type of maps from iteration space to address space that can be performed. The compiler gathers information about the supported iteration  spaces and map complexity from the available components and uses it to schedule and map operations, as described in \autoref{subsec:unified_buffer_mapping}.

The unified buffer abstraction, by creating a clear interface between the compiler and underlying hardware, gives the hardware architect the freedom to create a \textit{physical unified buffer} that implements this interface in a way that is both area and energy efficient.  To explore this hardware design space, we have created a flexible unified buffer hardware generator. The hardware generator creates both the logical design of the physical unified buffers and the compiler components that map a unified buffer abstraction to configuration registers in the hardware.

\subsection{Dual-Port SRAM}

\begin{figure}
    \centering
    \includegraphics[width=1.0\linewidth,keepaspectratio]{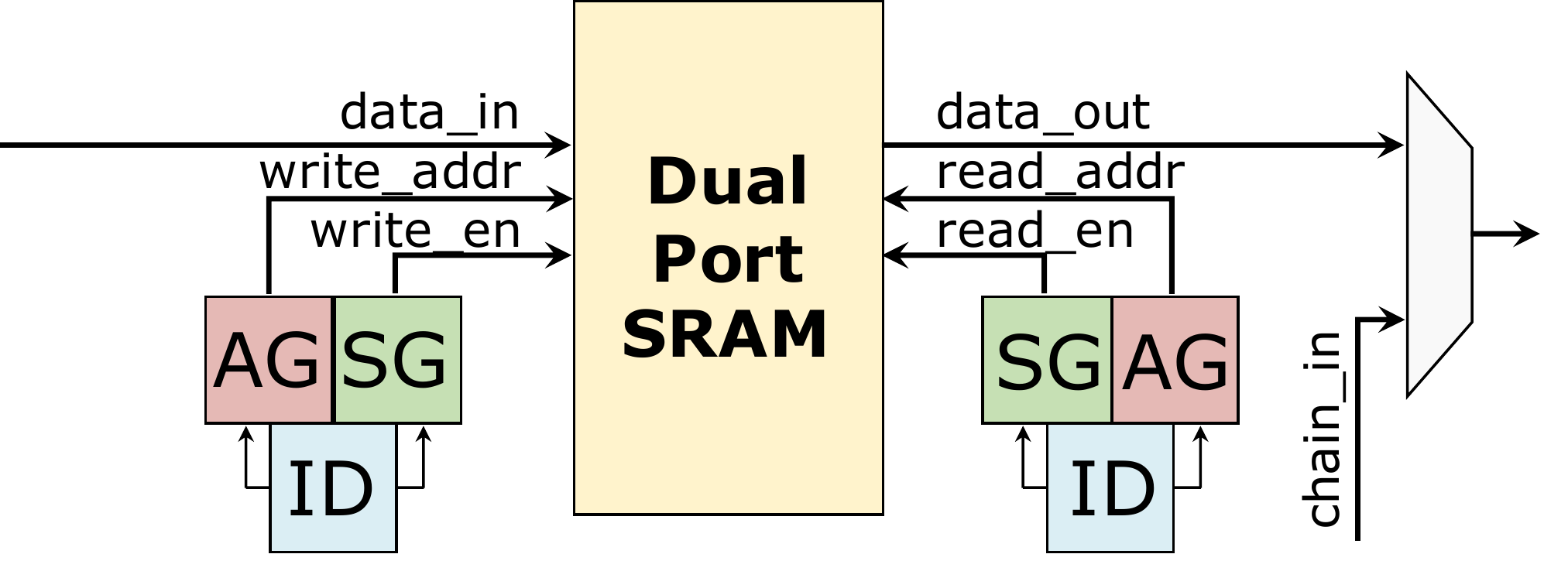}
    \caption{Simplified block diagram of a na\"ive physical unified buffer implementation with a dual-port SRAM. Two IterationDomain (ID) modules each drive an AddressGenerator (AG) and a ScheduleGenerator (SG) pair to orchestrate writes from and reads to the memory. There is a mux at the output to support memory chaining.}
    \label{fig:mem_tile_dual}
\end{figure}

The simplest hardware implementation
of a unified buffer wraps a dual-port SRAM with logic that computes the addresses and sequences of read/write enables for the iteration domain at each port, as shown in \autoref{fig:mem_tile_dual}. It also contains logic dedicated to chaining together multiple physical unified buffers into a larger buffer.

To support implementing the unified buffer abstraction, we instantiate three modules at each input and output port of the memory. These modules are IterationDomain (ID), AddressGenerator (AG), and ScheduleGenerator (SG), and they provide implementations of the corresponding components on the ports of the unified buffer abstraction. The IterationDomain module implements counters corresponding to a set of \texttt{for} loops, while the AddressGenerator and ScheduleGenerator modules implement the mapping logic from an IterationDomain module to an address and a read/write enable for the associated memory port. For the examples in this paper, we limit address maps and schedules to affine functions in keeping with the polyhedral model.

While dual-port SRAMs are the type of hardware that might be generated by a high-level synthesis (HLS) tool for an FPGA or an ASIC, they do not implement an efficient push memory for two reasons:  First, dual-port SRAMs are inefficient and can be over two times larger than their single-port counterparts for the same storage capacity while consuming 40\% more energy per access \cite{dualport_ratio_nautiyal}.  Second, energy per byte per access is often lower if more data is fetched from an SRAM on each cycle~\cite{vasilyev2019evaluating}. Thus, in custom-designed push memories, wide-fetch memories are typically used to emulate multiple ports and improve energy per access. Implementing wide-fetch memories requires two small additional memories that we describe next.

\subsection{Wide-Fetch Single-Port SRAM}
\label{sec:wide-fetch-pub}
To support a wide-fetch SRAM, we create a unified buffer that contains three unified buffers inside, where two of the buffers have small capacity---four to eight words per port when a four word fetch SRAM is used---and can be implemented using registers/register files. A buffer between the input port and the SRAM serves as a serial-to-parallel converter (aggregator (AGG)) and a buffer between the SRAM and the output port serves as a parallel-to-serial converter (transpose buffer (TB)). 
The software vectorization rules in \autoref{subsec:unified_buffer_mapping} provide transformations from the original set of iteration domains and access maps to a set of domains and access maps that treat each individual memory (the SRAM, the AGG, and TB register files) as separate physical unified buffers.

\begin{figure}
    \centering
    \includegraphics[width=\linewidth,keepaspectratio]{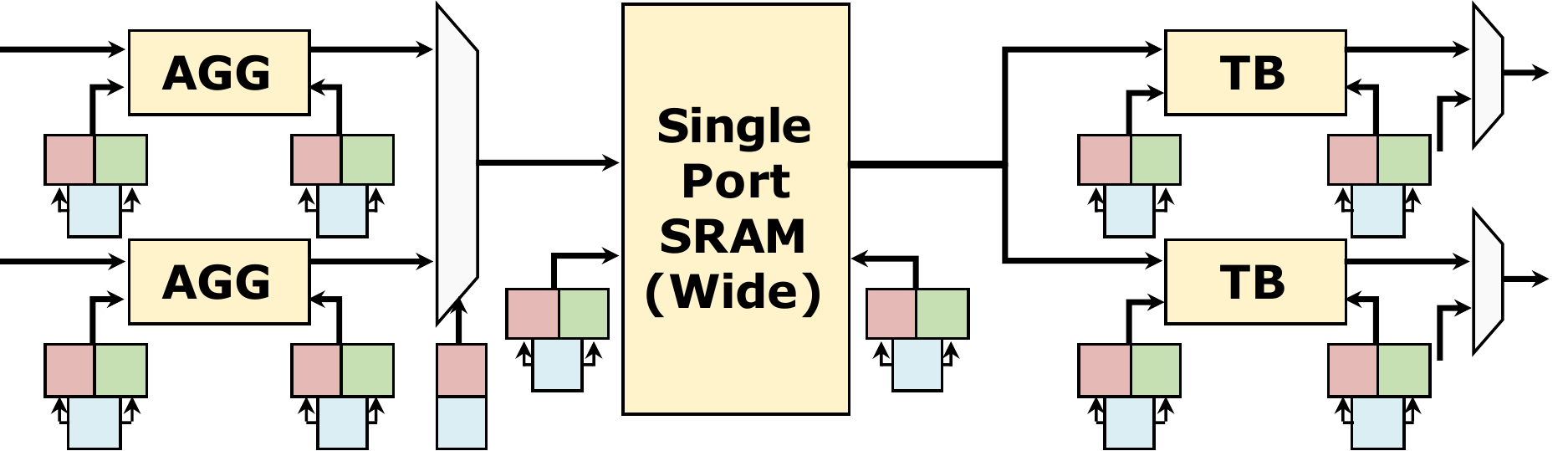}
    \caption{Simplified block diagram of a physical unified buffer with a wide-fetch single-port SRAM. Sets of ID/AG/SG controllers control the input and output of each memory. This implementation supports two input ports and two output ports by leveraging a 4 word fetch SRAM.}
    \label{fig:mem_tile_block_diagram}
\end{figure}

To support multiple input and output ports in a physical unified buffer implemented with a single-port SRAM, we need  some logic to allow for port sharing prescribed by the schedule. This support for port sharing is achieved by instantiating an ID and AG at the select line of a multiplexer that chooses which port accesses the SRAM at any given time. \autoref{fig:mem_tile_block_diagram} shows a high-level block diagram of the physical implementation of a push memory with 2 input ports and 2 output ports. Since the hardware implementation is only responsible for meeting the interface of a unified buffer abstraction, architects are free to optimize the design.

\begin{figure*}
    \centering
    \begin{subfigure}[t]{0.38\linewidth}
        \centering\includegraphics[width=0.95\linewidth]{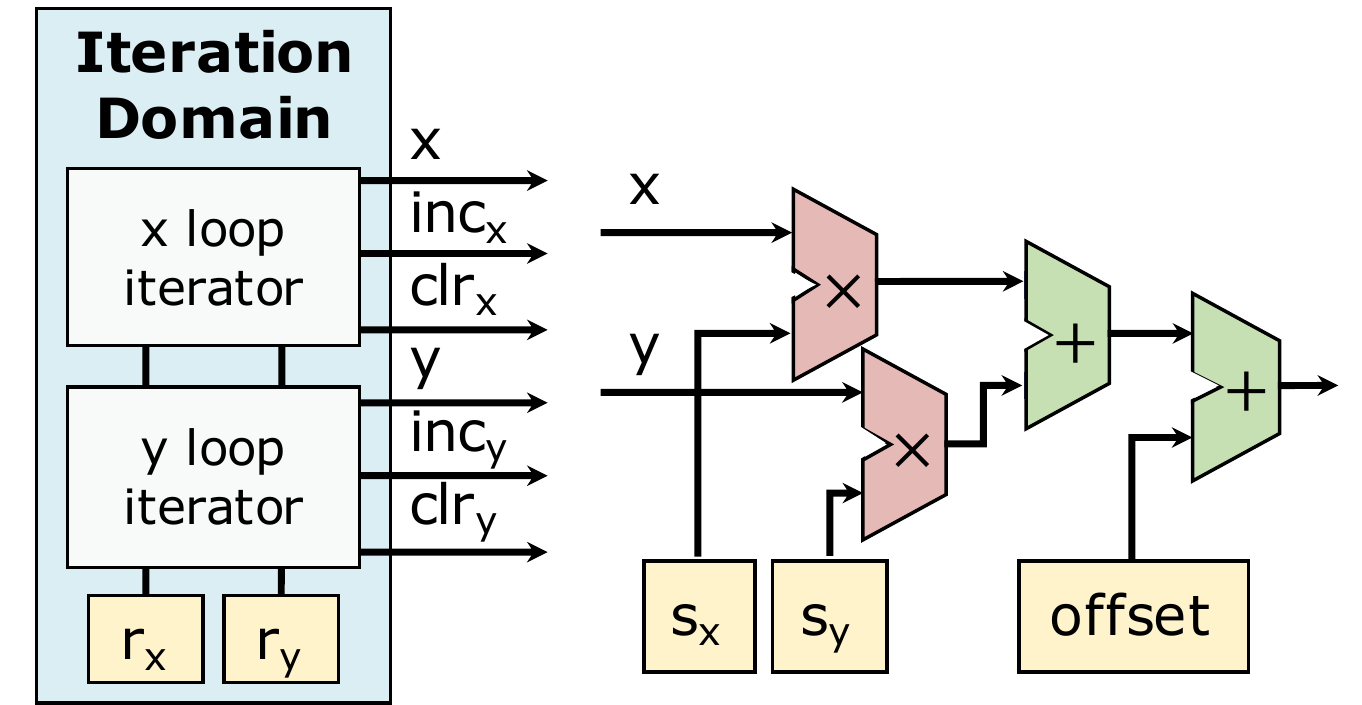}
        \caption{Basic affine function implementation}
        \label{fig:addr_gen_opts_a}
    \end{subfigure}
    \begin{subfigure}[t]{0.30\linewidth}
        \centering\includegraphics[width=0.95\linewidth]{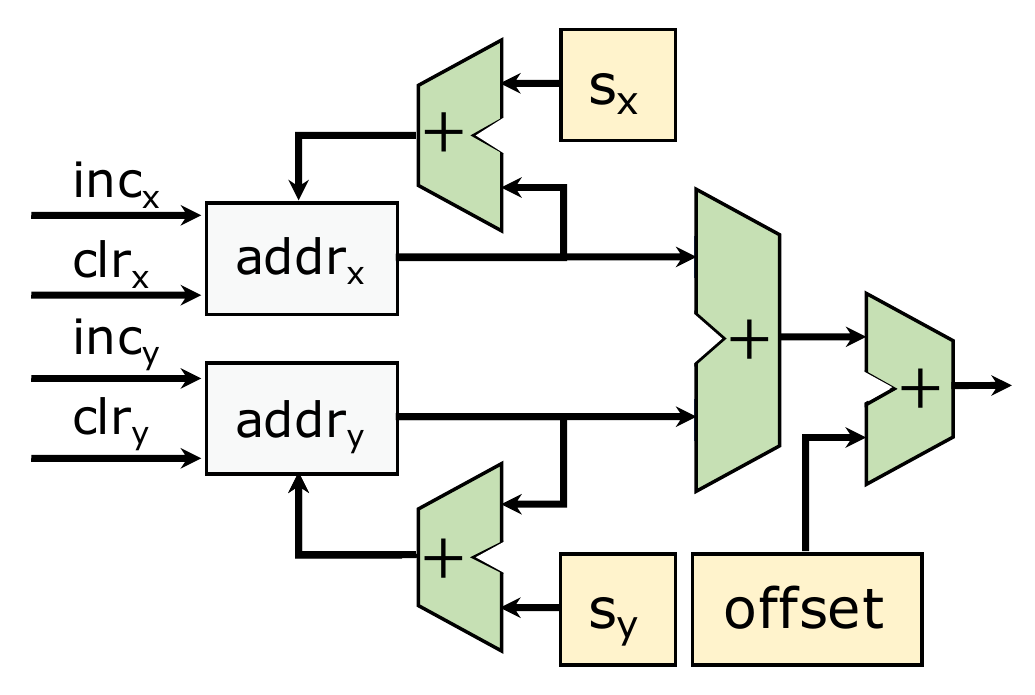}
        \caption{Affine function without multipliers}
        \label{fig:addr_gen_opts_b}
    \end{subfigure}
    \begin{subfigure}[t]{0.30\linewidth}
        \centering\includegraphics[width=0.95\linewidth]{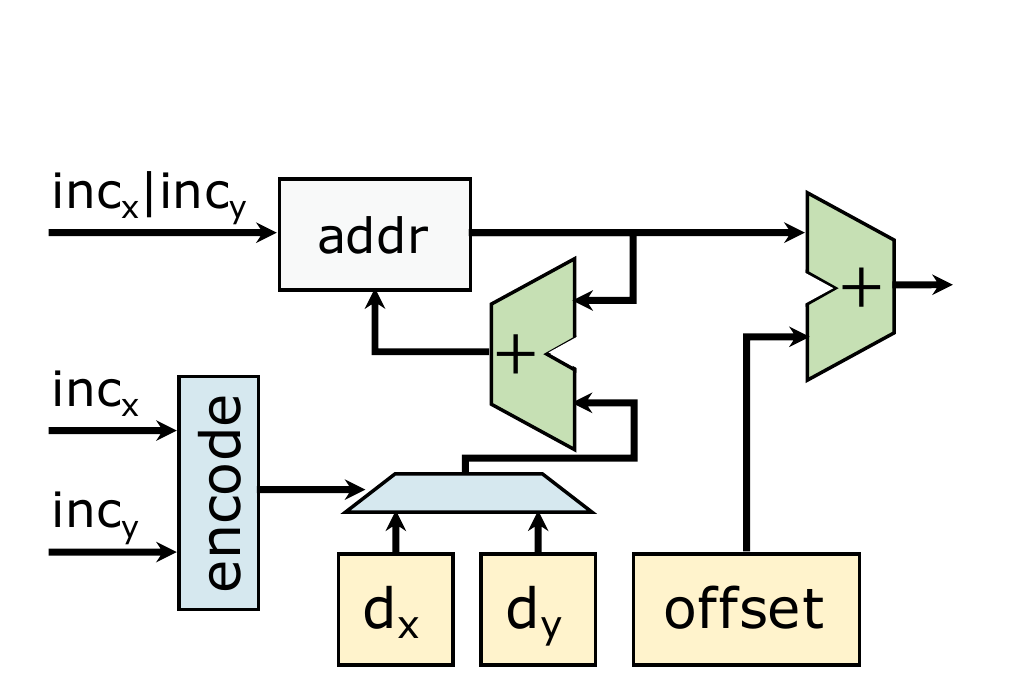}
        \caption{Affine function as recurrence relation}
        \label{fig:addr_gen_opts_c}
    \end{subfigure}
    
    \caption{Area optimizations in the affine function hardware for address and schedule generation with a two-dimensional iteration domain. \textbf{(a)} A basic implementation that uses the actual value of the counters in the iteration domain. \textbf{(b)} An implementation that keeps a running address contribution from each loop iterator without using its actual value. \textbf{(c)} The final optimized implementation that embeds the address delta between loop levels rather than the stride within each loop level.}
    \vspace{-6pt}
    \label{fig:addr_gen_opts}
\end{figure*}

\subsection{Optimizations}
\label{sec:physical-unified-buffer-optimizations}

We focus on two types of optimizations for creating efficient hardware implementations of unified buffers. These optimizations are topology-based resource sharing and the exploitation of recurrence in affine functions.

The topology-based resource sharing optimization comes from a key observation about unified buffers: sources and sinks have tightly coupled scheduling as any read from a memory ends in a write to a downstream memory in a statically determined number of cycles. In the case of our hardware design, we note that we only need one schedule generator to drive reads from the aggregator and subsequent writes to the SRAM. On the output side, this sharing is also possible, but a delay stage must be added between the schedule for SRAM reads and writes to the transpose buffer since the SRAMs we use have a delay of one cycle for reads. Figure \ref{fig:garnet} shows the resource sharing optimization applied on the buffer from Figure \ref{fig:mem_tile_block_diagram}.

\begin{figure}[h]
    \centering
    \includegraphics[width=\linewidth,keepaspectratio]{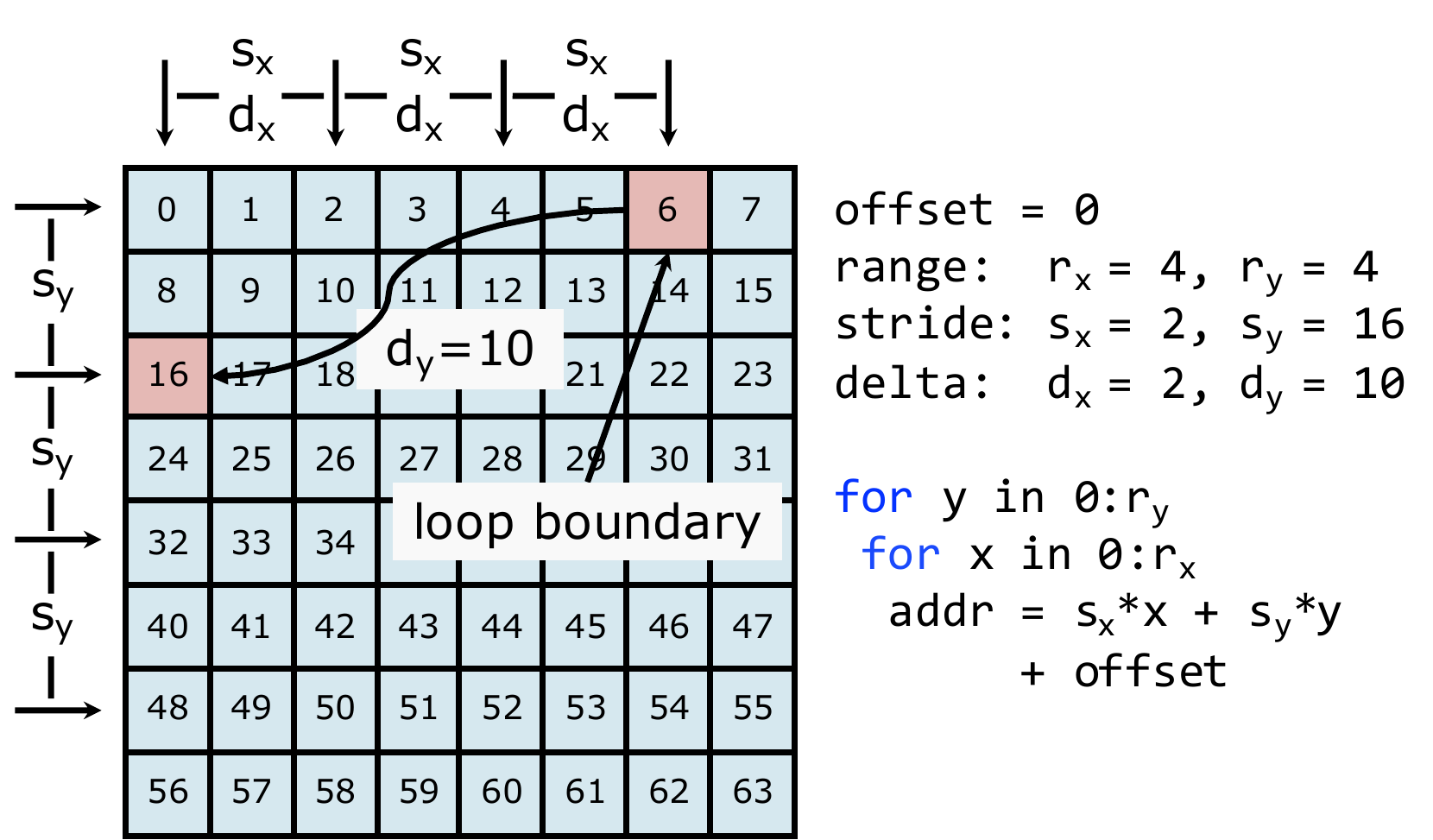}
    \caption{An example of a simple downsample-by-2 iteration pattern over an 8$\times$8 image. The relationship between the strides and deltas for a two-level loop nest are shown.}
    \label{fig:affine_example_down}
    \vspace{-6pt}
\end{figure}

The AddressGenerator and ScheduleGenerator modules can be described as affine functions of the loop iterators in the iteration domain. Figure \ref{fig:affine_example_down} shows an example of a two-dimensional affine function ($s_x * x + s_y * y + \text{offset}$). A na\"ive hardware implementation of such a two-dimensional affine function would use the design with two multipliers and two adders shown in \autoref{fig:addr_gen_opts_a}. This implementation explicitly computes the affine function of the raw loop iterator values $x$ and $y$. To eliminate the expensive multiplies, we can replace each multiplier with a register and an adder that simply increments the register by the configured stride when the respective iterator in the iteration domain increments. This optimization, with four adders and two registers, is shown in \autoref{fig:addr_gen_opts_b}. Note that while the implementation in \autoref{fig:addr_gen_opts_a} relies on the raw loop variables, the optimization in \autoref{fig:addr_gen_opts_b} only needs to know when the loop variables are incremented ($inc_x, inc_y$) or when they hit their boundaries in the iteration domain ($r_x, r_y$), denoted by $clr_x$ and $clr_y$. Accordingly, when $clr_x/clr_y$ is high, the respective register ($addr_x$, $addr_y$) is cleared along with the loop variable.

We can achieve another level of optimization by first noting that an explicit affine function of a set of loop variables can be formulated as a recurrence relation. Whereas the explicit formulation is: $$(x,y) \rightarrow s_x * x + s_y * y + \text{offset}$$ a recurrence relation can be set up as follows: $$(x, y)_{i+1} = (x, y)_{i} + (inc_y ? d_y : inc_x ? d_x : 0)$$ where $$(x, y)_{0} = \text{offset}$$ While $s_x, s_y$ represent multiplicative weights or strides for the respective loop variables, $d_x, d_y$ represent loop boundary deltas. If we assume that the loop variables are sorted as $x$ inside of $y$, then it is clear that $y$ will increment whenever the $x$ variable reaches its bound ($x = (r_x - 1)$) and wraps back around to 0 for the next iteration of the inner loop. In this case where the outer loop variable is incremented, we know that all inner loop variables are at their maximum bound, so we can define a delta $d_{outer}$ as a function of the respective strides and ranges of the inner loop variables as follows: $$d_{outer} = s_{outer} - \sum_{i=0}^{outer-1}{s_i * (r_i - 1)}$$ An example of the relation between the strides, ranges, and deltas is shown for a simple downsample-by-2 traversal of an 8$\times$8 image in \autoref{fig:affine_example_down}. This shift from an explicit to a recurrent representation of the affine function makes it easy to optimize the hardware implementation even further from \autoref{fig:addr_gen_opts_b}. Since we only need the delta for one loop variable at any time, we now only require a single adder and a register along with a multiplexer to increment the running address by the delta of the outermost loop variable that is incremented. This final hardware implementation is shown in \autoref{fig:addr_gen_opts_c}.

\section{Compiler Design}

Users of our system specify their applications in Halide, a high-level domain-specific language (DSL), to define computations and tiling structure. Halide separates the algorithm from the schedule to isolate computation from optimizations in execution~\cite{ragan2013halide}. The algorithm specifies the computation to get an output, while the schedule specifies the order in which the computation should be performed. 

Our compiler divides the problem of compiling the read-write buffers in Halide programs to push-buffer implementations into three steps: 
\begin{enumerate}
    \item The first step is the Halide scheduling system itself, whose scheduling language controls loop transformations and that we extend with accelerator commands.
    \item The second step uses polyhedral techniques to turn the multidimensional iteration spaces of Halide loops into one-dimensional cycle times at every buffer port, thus yielding pipeline parallelism. The same step then uses polyhedral techniques to extract the full specification of each buffer port in the unified buffer abstraction, as shown in \autoref{fig:ubuffer_spec}.
    \item The final step maps the abstract unified buffers to physical unified buffers built from low-level hardware primitives.
\end{enumerate}  

We chose to keep the Halide scheduling language for tiling instead of placing it in the second step (like the PLUTO scheduling algorithm \cite{bondhugula2008practical}).  The reason is that a high-quality general-purpose tiling algorithm for all dense linear algebra applications has not yet been found. As a result, we believe tiling is best left to either performance experts through a scheduling language or to domain-specific search procedures such as \cite{yang2020interstellar}. Thus, we limit our use of polyhedral techniques to memory analysis and semantic-preserving loop fusion.

\subsection{Halide Scheduling}

We extended the Halide scheduling language, which lets users define loop tiling but has no notion of push memories, to include a command to define what portion of an application should be placed on the accelerator. \autoref{fig:brighten_overview} shows an example. The placement is done by defining the accelerator output with \texttt{hw\_accelerate} and each of the accelerator inputs with \texttt{stream\_to\_accelerator}. After tiling a loop, the user can define which buffer variables should be defined as memories as opposed to fused with adjacent kernels by using \texttt{store\_at} and \texttt{compute\_at}, along the lines of \cite{pu2017programming}. Finally, the user specifies with \texttt{unroll} if some loops should be parallelized as opposed to run iteratively in several cycles. After these scheduling directives, all following optimizations and mapping are performed automatically without user input.

Additional simplifications are done in the frontend prior to buffer extraction. The frontend inlines constant arrays into the compute kernels to reduce the number of extracted buffers. This results in the mapped hardware using registers in the compute rather than instantiating these as memories. Another optimization is combining update statements, such as a series of adds in a reduction, to a single statement in order to reduce unneeded memory operations for our memory analysis. The Halide compiler then separates the Halide IR used for memories from the IR used for computation. The compute kernels are represented as a graph of operators and are used during the finishing steps.

\subsection{Unified Buffer Extraction}

\begin{figure}[t]
    \centering
    \includegraphics[width=1\linewidth,keepaspectratio]{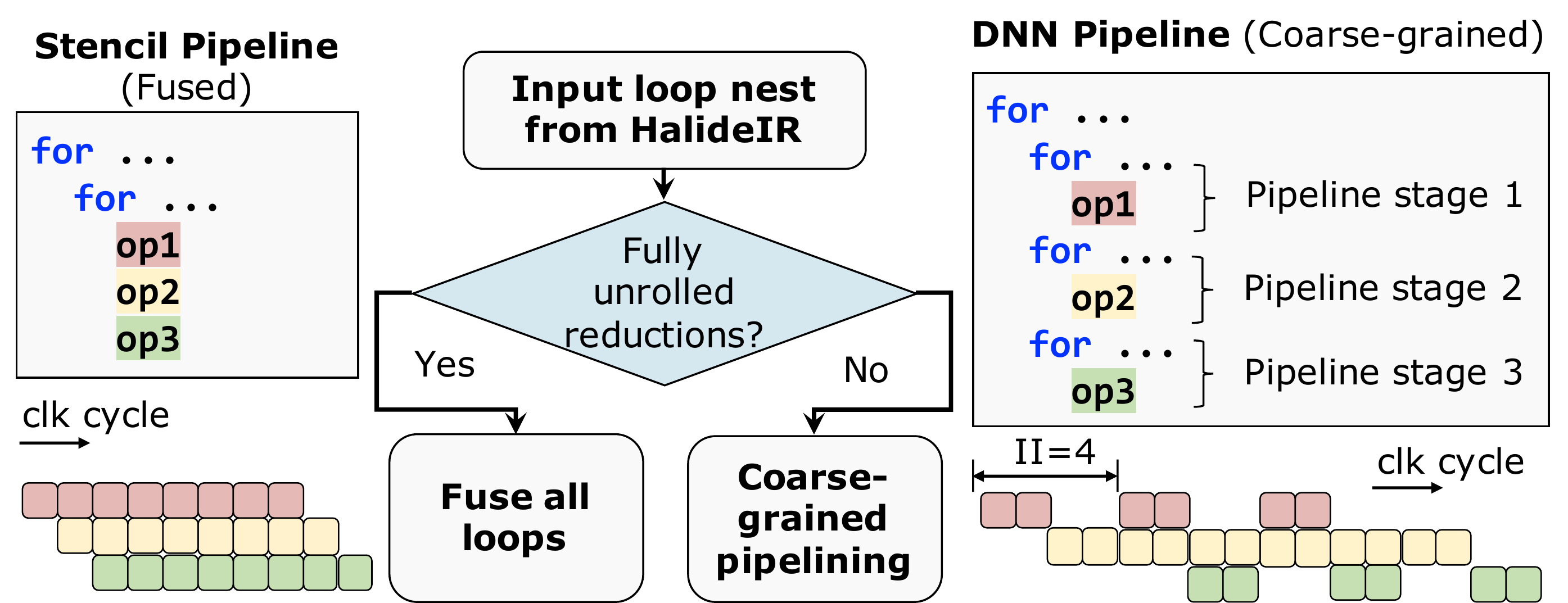}
    \caption{During unified buffer extraction, the compiler schedules pipelines based on the reduction loops. Above are two example pipelines: a stencil (left) and DNN (right).
    Each computation operation (op) corresponds to a buffer store and a set of loads.} 
\label{fig:ubuffer_extraction}
\end{figure}

The buffer extraction step analyzes the Halide IR to turn both loops and arrays in the Halide program
into push memories expressed using the unified buffer abstraction. That is, Halide programs describe computation as operations on arrays over iteration domains defined by index variables. To compile them to optimized push memories, buffer extraction analyzes buffer reads and writes to trace movement of values through memories. It then uses this information to distribute the
control flow across the address generators in push memories themselves.

Unified buffer extraction converts every Halide buffer into a unified buffer. Each memory reference to the Halide buffer is given a unique port
on the corresponding unified buffer. For each port, buffer extraction then computes an iteration domain, an access map, and a schedule. The
iteration domain is the Cartesian product of the bounds of the loops surrounding
the memory reference in the Halide IR, and the access map is the address expression of the corresponding memory reference.  The main work of unified buffer extraction is thus in defining
the cycle-accurate schedule that maps operations in the Halide program
to the cycle-times when they will happen in hardware.

Our cycle-accurate scheduler is designed to automatically exploit pipeline
parallelism in two broad classes of workloads: stencil pipelines from classical image processing and deep neural networks (DNNs). In classical image processing applications such as Harris
corner detection, the application consists of many stencil operations
that each produce output pixels from small windows of input pixels.
No one stage dominates the compute cost of the application
and every pixel in a given stage depends on a small
number of pixels in prior stages, making it easy to parallelize the execution
of producer and consumer stages. In DNN pipelines, on the other hand, a
single stage containing a large compute unit, typically a systolic array, dominates the
compute cost of the application, and pixels produced by that stage depend on large groups
of pixels from prior stages, making it difficult to parallelize across
stages.

Our scheduler detects and handles each of these pipeline
types separately. The scheduler selects the scheduling policy that will
be used with
a simple rule: If every reduction loop is fully unrolled, then it uses a scheduling strategy that is tailored to stencil pipelines,
and that produces a schedule that can be implemented efficiently
using line-buffers.
Otherwise, if there are any reduction loops that are not fully unrolled, it uses
an algorithm tailored to the DNN-style pipeline that uses coarse-grained
pipeline parallelism and double buffering to maximize utilization of
the most expensive compute unit as seen in~\autoref{fig:ubuffer_extraction}.
Both scheduling policies use the polyhedral analysis tool
ISL \cite{verdoolaege2010isl} to compute data dependencies between operations
and to solve the optimization problems used in formulating the schedule.

\textbf{DNN Pipeline} The DNN-style scheduler creates a schedule for a double-buffered pipeline. This pipeline is
coarse grained: operations on one tile of an image proceed
sequentially, but are overlapped with operations on the next
tile of the image. So, for example, while a convolution is being
computed on a tile that has already been loaded onto the CGRA, the
next tile is being loaded onto the CGRA. The first step in this scheduler is
to identify the tile loops that will be overlapped to
form the coarse-grained
pipeline. Our scheduler walks from the root of the program inward
and collects loop nests up to and including
the innermost loop of the application whose body is not a single perfect loop.
These perfectly nested loops form the outer coarse-grained pipeline.
We refer to the operations inside the pipeline as pipeline stages,
but it is important to note that these stages are themselves typically
loop nests. For instance, in the DNN pipeline pseudo code shown in ~\autoref{fig:ubuffer_extraction}, 
the outer coarse-grained pipeline is the for loop on line 1 and it contains three pipeline stages.

With the coarse-grained pipeline loops selected, the scheduler independently
creates a cycle-accurate schedule for each pipeline stage using
a standard HLS loop scheduler in the style of \cite{zhang2013sdc}.
It then starts the creation
of the coarse-grained pipeline by laying out each pipeline
stage sequentially, and setting
the initiation intervals (IIs) of the coarse-grained pipeline loops to values that
correspond to sequential execution.
Finally it reduces the IIs of the coarse-grained pipeline loops by binary searching
over their IIs until the compute unit of the largest
reduction stage is at 100\% utilization and all data dependencies are
respected. If the latency of operations in the DNN pipeline from 
~\autoref{fig:ubuffer_extraction} are 2, 4, 2 cycles respectively, 
the schedule will have coarse-grained pipeline $\text{II}=4$. 

\textbf{Stencil Pipeline} If the pipeline is classified as a stencil pipeline we apply
the scheduling algorithm described in \cite{huff2021clockwork}.
This algorithm produces a cycle accurate schedule in two stages.
First it fuses all loop nests in the application into a single
perfect loop nest. Then it computes a cycle accurate schedule for the
fused, perfect loop nest at an initiation interval of one.
The fusion is done incrementally, from the outermost
loop levels to the innermost. 
The fusion procedure uses an SDF-style constraint
problem to set the relative rates
and delays of each operation in a way that makes dependence distances as small
and uniform as possible.
Once fusion is finished, we compute a cycle accurate schedule
for the loop nest using the same HLS scheduler that
is used for the pipeline stages of the double-buffered pipeline.

With scheduling finished,
all operations have been assigned to clock cycles (one-dimensional affine schedules)
and the bandwidth requirement of each memory is known. Now the
task of the compiler is to synthesize the abstract unified buffers
into buffer implementations built out of the available physical primitives.

\begin{figure}
    \centering
    \begin{subfigure}[t]{0.22\linewidth}
        \captionsetup{font={small}}
        \centering\includegraphics[width=\linewidth]{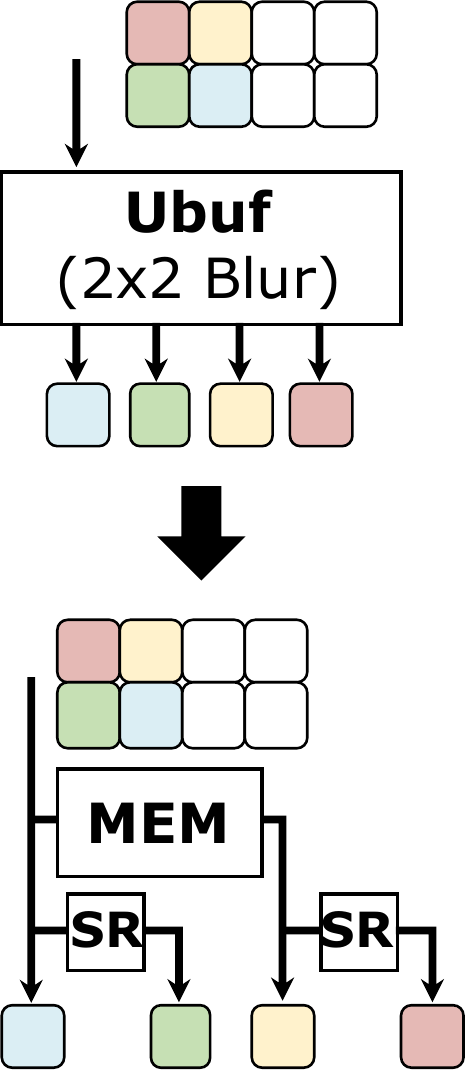}
        \caption{Shift register optimization}
        \label{fig:buffermap_a}
    \end{subfigure}
    \begin{subfigure}[t]{0.25\linewidth}
        \captionsetup{font={small}}
        \centering\includegraphics[width=\linewidth]{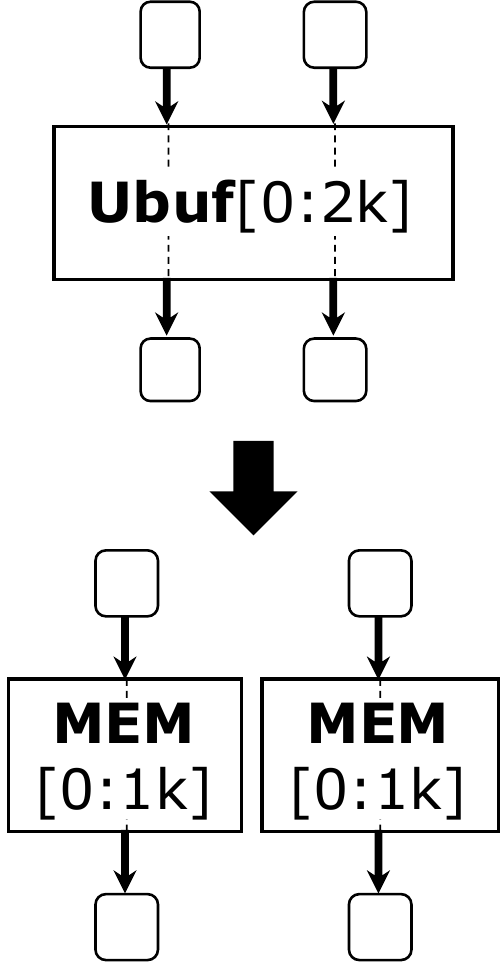}
        \caption{Banking for higher band-width}
        \label{fig:buffermap_b}
    \end{subfigure}
    \begin{subfigure}[t]{0.25\linewidth}
        \captionsetup{font={small}}
        \centering\includegraphics[width=\linewidth]{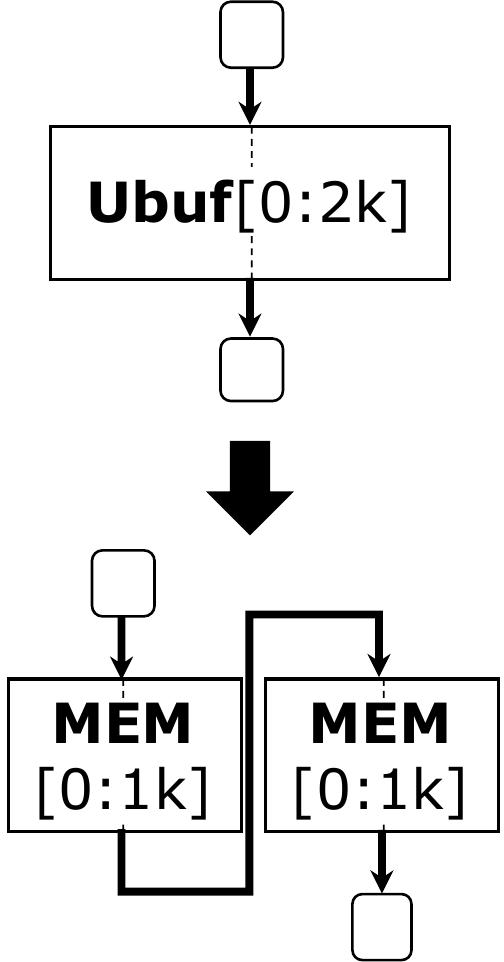}
        \caption{Chaining for higher capacity}
        \label{fig:buffermap_c}
    \end{subfigure}
    \begin{subfigure}[t]{0.19\linewidth}
        \captionsetup{font={small}}
        \centering\includegraphics[width=\linewidth]{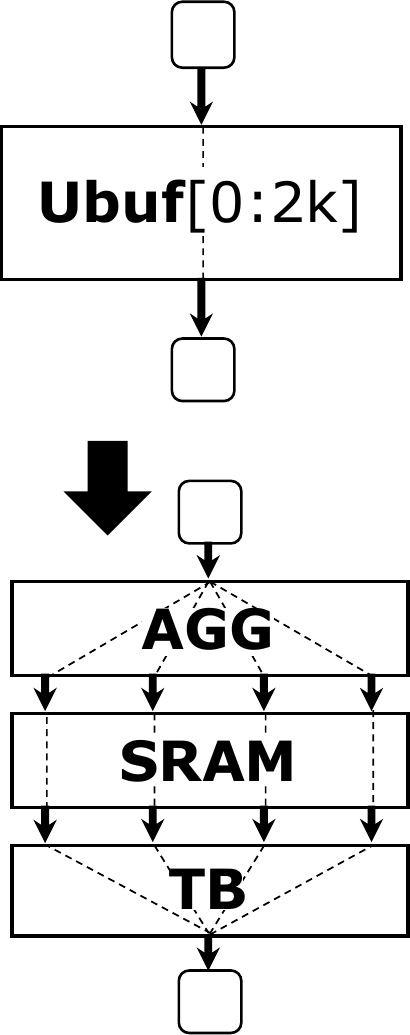}
        \caption{Vectorization for wider fetch}
        \label{fig:buffermap_d}
    \end{subfigure}
    \caption{Transformations applied to map an abstract unified buffer to physical unified buffers (SR: Shift Register, AGG: Aggregator, TB: Transpose Buffer).}
    \label{fig:buffermap}
\end{figure}

\subsection{Unified Buffer Mapping}
\label{subsec:unified_buffer_mapping}

The next step in
compilation is to map the unified buffers and compute kernels in the application graph
to physical unified buffers and compute units on the accelerator. This mapping
produces the configuration bits for each
physical unified buffer and compute unit used in
the design.

In principle, the unified buffers extracted from the Halide IR can be mapped directly to physical unified buffers on the target accelerator.  In practice, however, this is rarely possible for the following reasons:

\begin{itemize}
\item \textbf{Physical unified buffer bandwidth} The physical unified buffers
    on the accelerator may not have sufficient bandwidth. For example, one version of our accelerator only has a
    single 4 word-wide SRAM in each physical unified buffer meaning that,
    even with balanced numbers of reads and
    writes, each buffer can only support up
    to 4 memory operations per cycle.
    However, unified buffers such as the \textit{brighten} buffer from our
    example need to perform 5 memory operations
    per cycle, and many common image processing
    patterns, such as the $3\times3$ convolution, need
    9 reads and 1 write per cycle in their
    steady state if the convolution kernel is unrolled.
    \item \textbf{Wide fetch width} Accesses in Halide
    programs may have any integer bit width, but accesses in physical unified buffers can have a vectorized fixed width
    that is wider than the individual pixels in
    a Halide application. For example, 
    accesses to the 4 word-wide SRAM in the physical unified buffers we built are done in vectors
    of four 16 bit integers, with 4 word-wide data vectors
    buffered in the aggregator and the transpose
    buffer between writes and
    reads to the SRAM.
    \item \textbf{Multidimensional address} Physical memory is linear with a single dimension, 
    but our unified buffer abstraction can support arbitrary data dimensions.
    \item \textbf{High capacity} The cycle-accurate scheduler 
    reduces storage requirements by
    improving locality, but
    even after storage minimization,
    unified buffers may need more space than what is
    available in a single physical unified buffer.
\end{itemize}

\textbf{Shift Register Optimization and Banking} To address the
need for high bandwidth, each unified buffer must be broken down
into smaller unified buffers that can each be mapped to a single physical unified buffer. Our compiler has two strategies for servicing
high bandwidth accesses: shift register
introduction and banking. Shift register
introduction is possible whenever the
dependence distance between one port
(the source) and another (the destination)
is constant, and the set of values that
appear on the source is a superset of
the values that appear on the destination.
Our compiler performs an exhaustive shift register
analysis that finds all opportunities
to convert output ports into shift
registers fed from other ports (\autoref{fig:buffermap_a}).
For instance, according to \autoref{fig:ubuffer_spec}, the buffer feeding the $2\times2$ blur kernel has four output ports, whose dependence distances to the input port are 0, 1, 64, 65 respectively. As shown in \autoref{fig:buffermap}a, this example can be implemented with two shift registers and a memory that delays by 64 cycles.

After shift register introduction is
complete, any remaining ports must be serviced from
banks of
memory with address generators (\autoref{fig:buffermap}b). Our compiler uses
a simplified version of an optimal banking algorithm
for stencil computations \cite{escobedo2018graph} to
find legal banking schemes for the remaining ports.

\textbf{Vectorization} To make
efficient use of physical unified buffers with wide-fetch SRAMs, the access patterns
of the buffers must be broken
into sub-sequences with the same length as the SRAM fetch width. At each input port of the buffer, this sub-sequence is assembled serially by the aggregator (AGG). Once the aggregator is full, the sub-sequence is written to the SRAM. At each output port, when the transpose buffer (TB) is empty, it receives a sub-sequence from the SRAM, which it then sends out serially on the output port. %

\begin{figure}[t]
    \centering
    \includegraphics[width=1\linewidth,keepaspectratio]{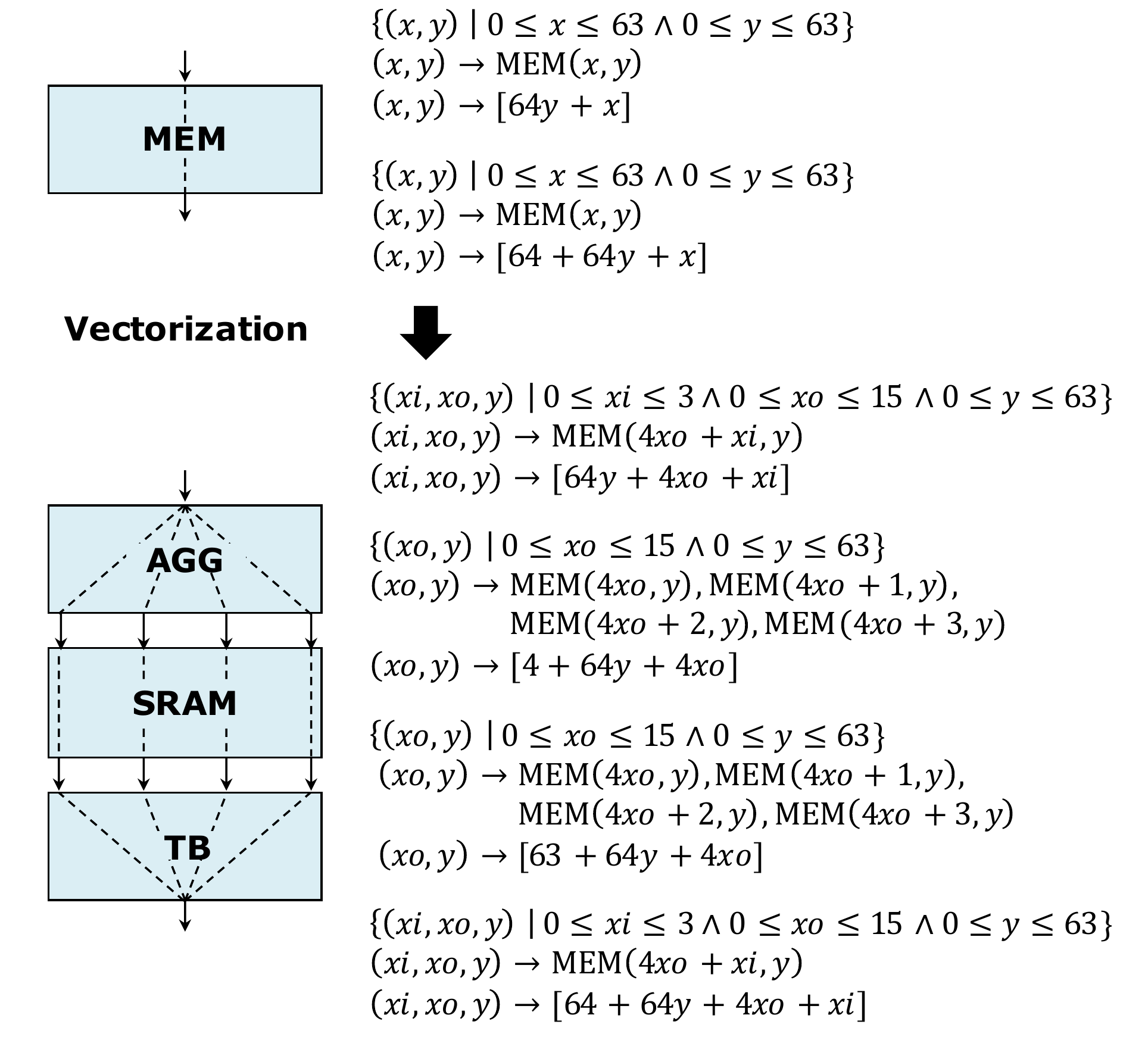}
    \caption{An abstract unified buffer before and after vectorization. The unified buffer \textbf{MEM} is the memory in the brighten and blur application after shift register introduction from \autoref{fig:buffermap_a}.} %
\label{fig:vectorization}
\end{figure}

We can think of the introduction of the AGG, SRAM, and TB components as strip-mining the innermost loops of the original
program and introducing wide fetch-width loads and stores to these components.
As an example, \autoref{fig:vectorization} shows a unified buffer where the buffer \texttt{MEM} adds a 64 cycle delay between input and output streams. 
It has a two-dimensional iteration domain with index variables $x$ and $y$. We apply the following transformation to strip-mine its iteration domain:
\begin{align}
     (x, y) \rightarrow & (x\bmod\textsc{fw}, \texttt{floor}(x/\textsc{fw}), y) \label{eq:vec_trans}
\end{align}
where $\textsc{fw}$ is the fetch width of the wide-fetch SRAM. This transformation creates a third dimension in the iteration domain for data aggregation and data serialization in AGG and TB respectively. 

Next, we apply the following iteration domain transformation on the ports of the SRAM to allow wide writes and reads:
\begin{align}
    (x, y) \rightarrow &
    (\texttt{floor}(x/\textsc{fw}), y) \label{eq:vec_trans_para}
\end{align}
As shown in \autoref{fig:vectorization}, our compiler automatically creates the access maps and schedules at the SRAM ports and records them in the abstract unified buffer. It also adjusts the schedules of aggregator to SRAM and SRAM to transpose buffer transactions to minimize the storage requirement in AGG and TB while respecting all data dependencies. Finally, it maps the rewritten access map and schedule to the address generation and scheduling hardware shown in \autoref{fig:mem_tile_block_diagram}.  

\textbf{Address Linearization}
 The access pattern in a unified buffer supports an arbitrary number of data dimensions, but the physical unified buffer requires the N-dimensional addresses to be converted to a single dimension. So, first, as shown in the following equation, an inner product is applied between each N-dimensional address $\vec{a}$ and an offset vector $\vec{o}$, which encodes the memory layout. 
 \begin{equation}
    \{ \text{MEM}[a_0, a_1, ..., a_{N-1}] \rightarrow \text{MEM}[\Sigma^{i} a_{i}  \cdot o_i]\} \label{eq:linear_addr}
\end{equation}
 Consider the memory in the previous section as an example. The blur kernel's input image size is $64 \times 64$. Polyhedral analysis identifies that there are a maximum of 64 live pixels. Storage minimization infers that a circular buffer must be implemented, so the compiler calculates the inner product of $\{x, y\}$ and the offset vector $\{1, 64\}\bmod 64 = \{1, 0\}$, which results in the linear address $x\times 1 + y\times 0 = x$.
 
\begin{figure}
    \centering
    \includegraphics[width=1\linewidth,keepaspectratio]{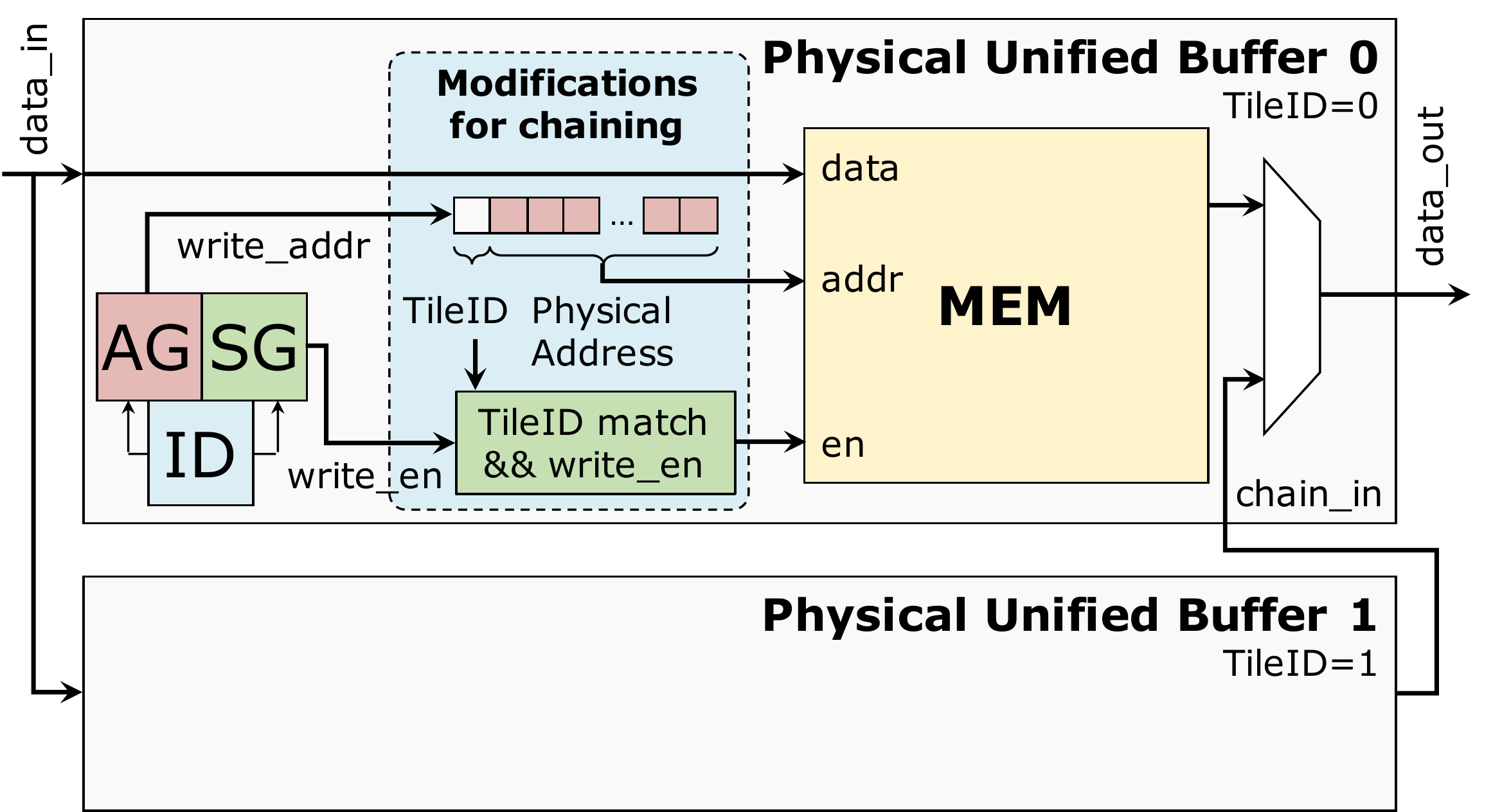}
    \caption{Architectural modifications to the buffer from Figure \ref{fig:mem_tile_dual} to support chaining multiple physical unified buffers. The figure shows the changes for writing to the memory, similar modifications exist for reading.} %
    \label{fig:chaining_hw}
\end{figure}

\textbf{Chaining} 
To map unified buffers with higher capacity than that of one physical unified buffer, we use a strategy called chaining to couple several buffers into a single logical buffer (\autoref{fig:buffermap}c). 
The hardware implementation of chaining is depicted in \autoref{fig:chaining_hw}.
In this implementation, each memory tile on the CGRA is assigned a unique tile ID.
Our compiler statically analyzes the access map and the schedule of the unified buffer, and partitions the access map into pieces to be implemented by separate physical buffers chained together. \autoref{eq:tileid} and \autoref{eq:chaining} transform a logical address, $a$, in the access map into a tile ID and a physical address in the memory tile, using the capacity $C$ of the memory tile. 
\begin{align}
    \{ \text{MEM}[a] \rightarrow& \text{TileID}[\texttt{floor}(a/C)]\label{eq:tileid} \\
    \text{MEM}[a] \rightarrow& \text{PhysicalAddress}[a\bmod C] \}
    \label{eq:chaining}
\end{align}
For example, assume a SRAM macro with a capacity of 32 words (unrealistic, but used for demonstrating chaining). Our delay buffer from the brighten and blur application is then implemented by chaining two memories. Thus $\text{TileID}(x,y) = floor(x/32)$ and the physical address is $\text{PhysicalAddress}(x,y) = x \bmod 32$. This expression is later mapped down to configurations in the AG and SG shown in \autoref{fig:addr_gen_opts}. 

\textbf{Finishing Steps} Finally, we map the compute kernels produced by the Halide frontend to processing elements (PEs) on the CGRA. We place and route (PnR) this mapped graph of PEs  and physical unified buffers on the CGRA following standard multi-stage optimization with global PnR followed by detailed PnR to obtain the final configuration bitstream.

\begin{figure}
    \centering
    \includegraphics[width=1.0\linewidth,keepaspectratio]{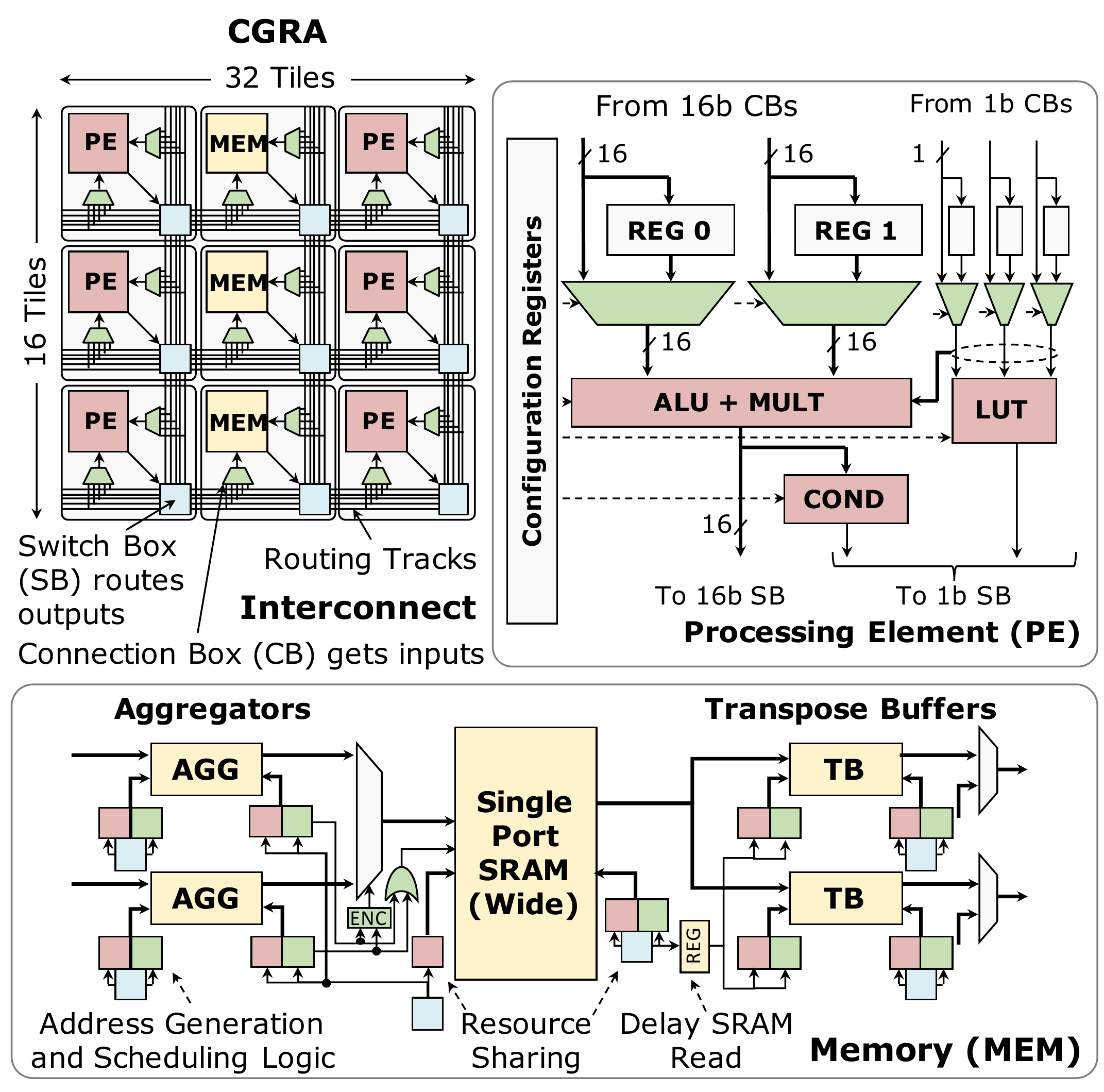}
    \caption{The architecture of our CGRA is a $16\times 32$ array of processing element (PE) and memory (MEM) tiles. One-fourth of the tiles are MEMs and the rest are PEs. The memory tile contains the optimized physical unified buffer described in Section \ref{sec:physical-unified-buffer-optimizations}.}
    \label{fig:garnet}
\end{figure}

\section{Evaluation}
To evaluate our compiler, we use it to map applications to 
both a Zynq UltraScale+ 7EV FPGA and a coarse-grained reconfigurable array (CGRA). When targeting the FPGA, our compiler outputs synthesizable C that is fed into Xilinx's Vivado system which synthesizes, places, and routes the resulting design at 200 MHz.  We include Vivado's report of the resources, performance, and energy.%

\begin{figure}
    \centering
    \includegraphics[width=1.0\linewidth,keepaspectratio]{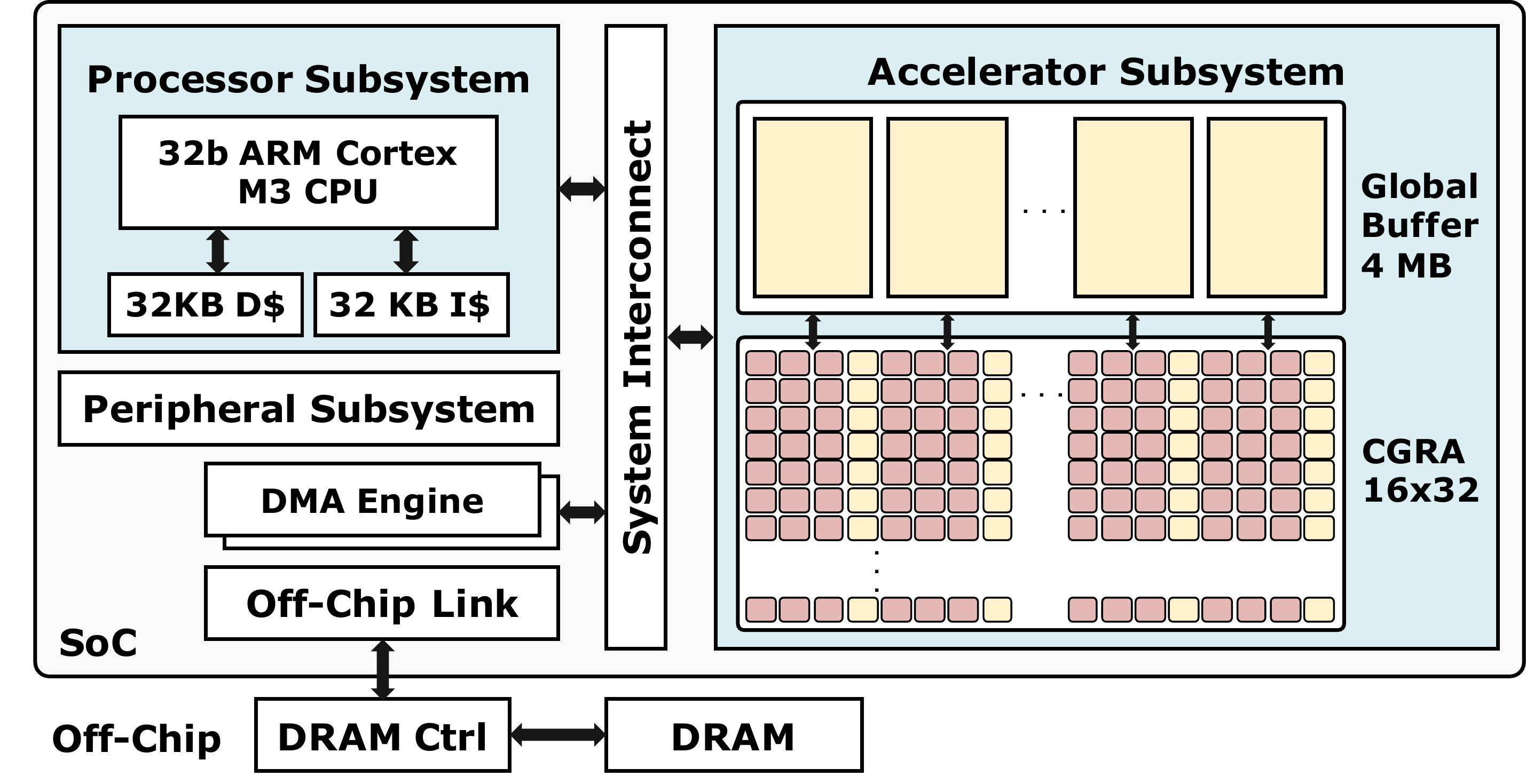}
    \caption{SoC consisting of the CGRA and the global buffer, which provides a deterministic access latency to the CGRA.}
    \label{fig:soc}
\end{figure}

The CGRA, shown in \autoref{fig:garnet}, resembles an island-style FPGA, with LUTs replaced by processing element (PE) tiles with 16 bit integer ALUs and BRAMs replaced by memory (MEM) tiles containing physical unified buffers. As shown in Figure~\ref{fig:soc}, the CGRA is a part of a full system-on-chip (SoC) where it connects to a large multi-banked, double-buffered memory called the global buffer. The data tiles required by the CGRA are first brought into the global buffer and then are streamed into the CGRA from there. This allows computation on the current tile in the CGRA to be overlapped with the movement of the next tile into the global buffer. The global buffer provides a deterministic access latency to the CGRA and hides the non-deterministic latency of the main memory. If the computation on the current tile completes before the next tile is brought into the global buffer, the whole CGRA is stalled until the data tile becomes available. Such coarse-grained stalling does not hamper the performance of compute-dominated image processing and machine learning applications and allows the compiler to perform fully static scheduling on the CGRA.

When targeting the CGRA, our compiler outputs a logical description of the design which uses custom mapping, place and route tools designed for this CGRA. Once an application has been created in Halide, all further steps happen automatically without manual annotation. To generate power and area numbers, we created a complete design of this CGRA and used Cadence Genus and Innovus tools to synthesize and place and route the physical unified buffers and processing elements of the CGRA in TSMC 16nm technology. We then extracted power and area numbers from this completed design. 
\begin{table}[t]
   \caption{
        Three different implementations of a physical unified buffer (UB): a dual-port (DP) SRAM with PEs for address generation (AG), a DP SRAM with optimized AG, and our final physical UB with a single-port (SP) SRAM with fetch width of 4, aggregator, and transpose buffer each with their AGs. The total area and energy needed for a $3\times3$ convolution decreases as we specialize the physical unified buffer.
    \label{tab:mem-tile-area}
    }
    \centering
    \begin{tabular}{p{0.36\columnwidth}C{0.07\columnwidth}C{0.12\columnwidth}C{0.09\columnwidth}C{0.1\columnwidth}} \toprule
         &  MEM Area ($\mu m^2$)   & SRAM Area (\%)     & Total UB Area   & Energy (pJ / access) \\ \midrule
         DP SRAM + PEs (Baseline)        & 19k   & 82   & 34k & 4.8 \\ \midrule
         DP SRAM + AG                    & 23k   & 70   & 23k & 3.6 \\ \midrule
         4 Wide SP SRAM + AGG + TB + AGs & 17k   & 32   & 17k & 2.5 \\ \bottomrule
    \end{tabular}
\end{table}

\subsection{Benefit of the Physical Unified Buffer Primitive}

A physical unified buffer includes address generation and control logic in addition to the SRAM storage array. While this logic adds area, it is far more efficient than using the processing elements (PEs) on a CGRA for addressing and control as shown in the second row of \autoref{tab:mem-tile-area}.  Adding this logic to a dual-port $2048\times16$ bit SRAM (\autoref{fig:mem_tile_dual}) does decrease the array efficiency of the memory tile, but reduces the area and energy by 32\% and 25\% respectively, compared to the version where the addressing and control is implemented on PEs (baseline).

Further improvements are possible by removing the need for dual-port SRAMs. The area of the dual-port $2048\times16$-bit SRAM is around $2.5\times$ larger than the single-port $512\times64$-bit SRAM.  Thus, as the third row of \autoref{tab:mem-tile-area} shows, even though the array efficiency of using a single-port SRAM with extra aggregation and transpose logic as described in \autoref{sec:wide-fetch-pub} is only half of the prior versions, it again results in a buffer that is a 26\% smaller and 30\% lower energy than the best dual-ported version.  Integrating the address generation and using efficient SRAM macros yields a buffer that is half the area and energy of the original design, and leaves our unified buffer with energy costs that are dominated by the fetch energy of the underlying SRAMs.

\begin{table}[t]
    \caption{
        Halide applications used in the evaluation section.
        \label{tab:apps}
    }
    \centering
    \begin{tabular}{p{0.13\columnwidth}p{0.09\columnwidth}p{0.63\columnwidth}} \toprule
         Application  & Type    & Description \\ \midrule
         gaussian     & stencil & $3\times3$ convolutional blur \\
         harris       & stencil & corner detector using gradient kernels \\
         upsample     & stencil & up sampling by repeating pixels \\
         unsharp      & stencil & mask to sharpen the image \\
         camera       & stencil & demosaicing and image correction \\
         resnet       & DNN     & layer using multi-channel convolution\\
         mobilenet    & DNN     & layer using separable, multi-channel convolution\\ \bottomrule
    \end{tabular}
\end{table}

\subsection{Performance, Power, and Utilization for Applications}

\begin{figure}
    \centering
    \includegraphics[width=0.9\linewidth,keepaspectratio]{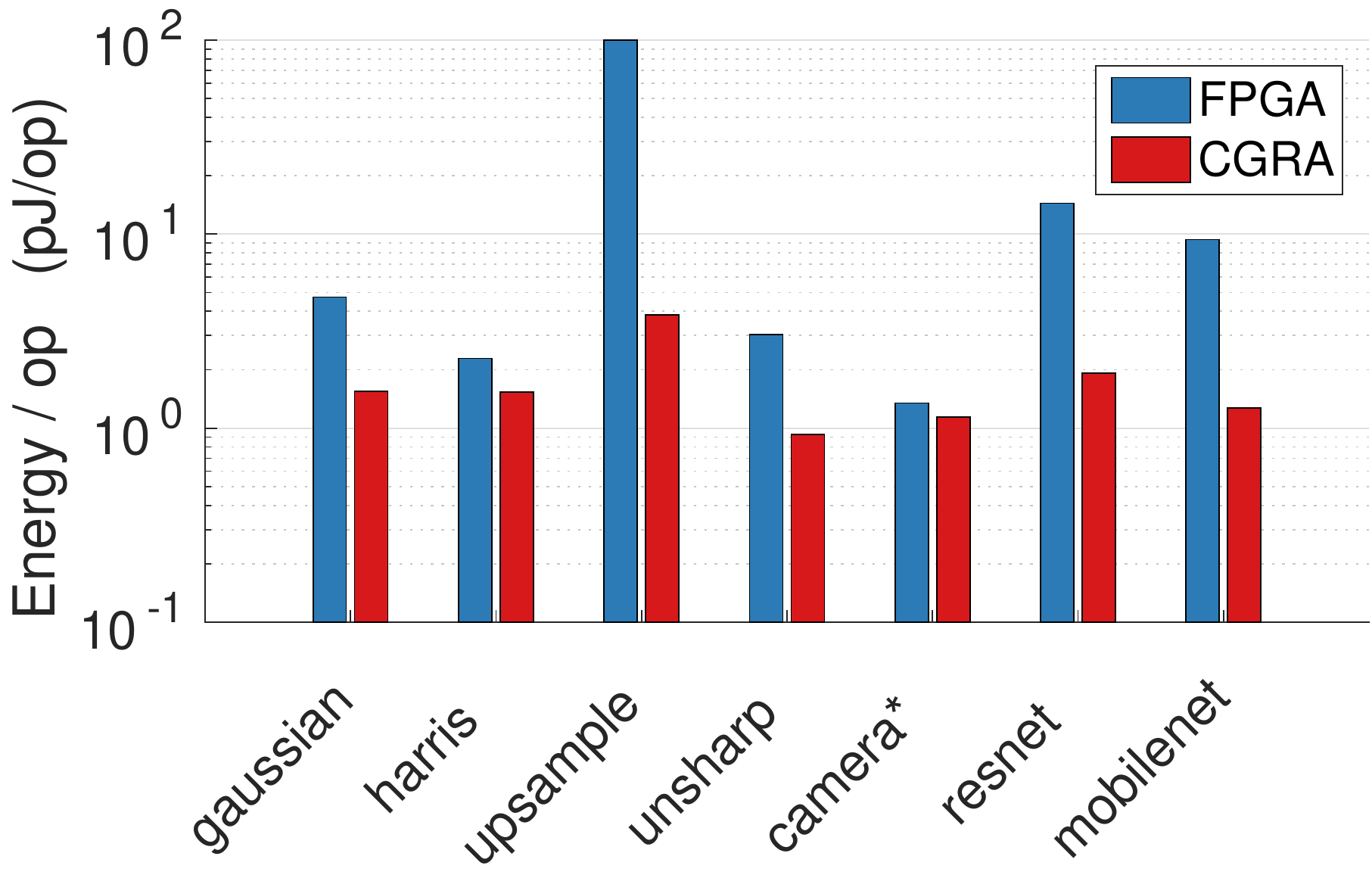}
    \caption{Comparison of the energy per operation (op) for running applications on a CGRA vs FPGA.\protect\footnotemark}
    \label{fig:energy}
\end{figure}

We use our compiler to compile the applications listed in \autoref{tab:apps} to both FPGA and CGRA targets. 
The compiler frontend transforms each Halide application to a polyhedral IR.
From this IR, our compiler can generate FPGA and CGRA bitstreams. To generate an FPGA bitstream, it generates synthesizable C code that feeds into Vivado HLS to produce full rate designs (II=1). To generate a CGRA bitstream, it uses the CGRA backend from \autoref{subsec:unified_buffer_mapping}.  
\autoref{tab:fpga-results} shows the resource utilization for applications mapped to an FPGA and CGRA using our framework. Since our results do not depend on the size of the application, to minimize time spent running the power analysis tools, we used smaller problems sizes, which leads to the modest number of resources shown. \autoref{fig:energy} shows the resulting energy/op consumed on both the FPGA and the CGRA. The more efficient unified buffer implementation and optimized 16 bit logic mean that the CGRA is $4.3\times$ more efficient than the FPGA.

\begin{figure}[t]
    \centering
    \includegraphics[width=0.9\linewidth,keepaspectratio]{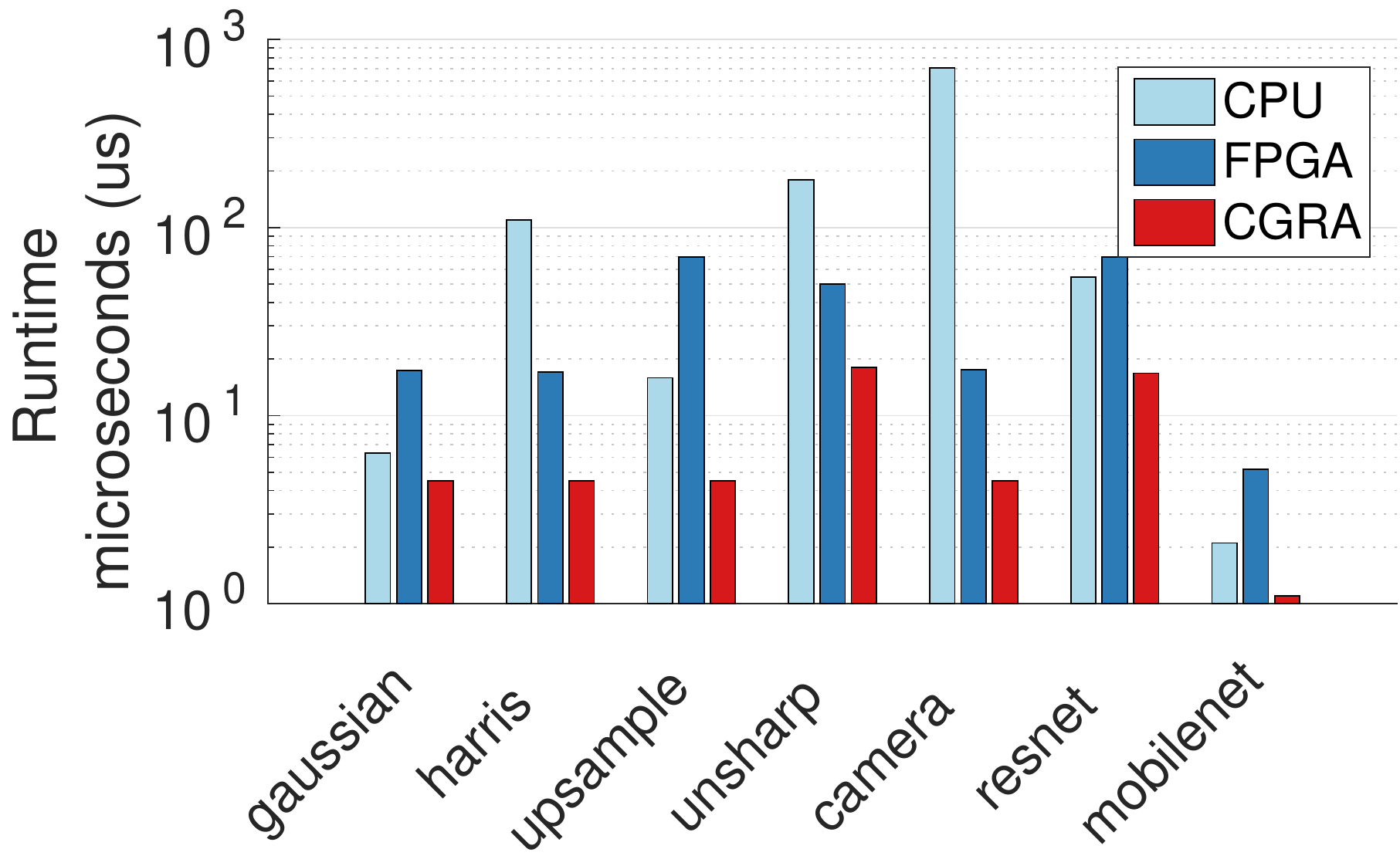}
    \caption{A comparison of application runtimes on CGRA, FPGA, and CPU.}
    \label{fig:runtime}
\end{figure}

\autoref{fig:runtime} shows the applications' runtime on the CGRA, FPGA, and a CPU. Our CPU comparison is an Intel Xeon 4214 with 16.5 MB cache with a 2.2 GHz base frequency. We use the same Halide application code for each backend, and then validate the output images against each other. Even though we are using modest hardware resources, the CGRA is able to outperform the Intel CPU. The CGRA dominates the FPGA due to its higher clock frequency (900 MHz). 

\begin{table}[t]
    \caption{
        Area results for various applications on FPGA and CGRA.
        \label{tab:fpga-results}
    }
    \centering
    \begin{tabular}{p{1.2cm} p{0.6cm} p{0.4cm} p{0.6cm} p{0.8cm} |p{0.7cm} p{0.7cm}} \toprule
         & \multicolumn{4}{c|}{FPGA Usage (\#)}       & \multicolumn{2}{c}{CGRA Usage (\#)}         \\
                              & BRAM   & DSP  & FF   & LUT   & PEs  & MEMs   \\ \midrule
         gaussian             & 0      & 1    & 437  & 863   & 19   & 1  \\
         harris               & 0      & 2    & 2449 & 4138  & 83   & 5  \\
         upsample             & 1      & 3    & 848  & 1923  & 0    & 1  \\
         unsharp              & 8      & 6    & 1954 & 2784  & 56   & 6  \\
         camera               & 5      & 7    & 1542 & 4448  & 397  & 8  \\
         resnet               & 16     & 64   &11100 & 6957  & 128  & 81  \\
         mobilenet            & 0      & 48   & 5100 & 5692  & 114  & 7 \\ \bottomrule
    \end{tabular}
\end{table}

\footnotetext{The camera application does not fit on our CGRA, so we estimated its power based on the PE power consumption in our other stencil applications.}

\subsection{Application Schedule Exploration}

Using Halide, we can modify the schedule to trade throughput for area to meet different design requirements. For example, by using Halide's \texttt{store\_at} primitive, we can choose which buffers should be created in an application. We show
the resource consumption and runtime for six
different schedules of Harris in \autoref{tab:sch-exploration} after running through the compiler (ending before the mapping stage). The first three schedules show the effect of buffering versus recomputing
intermediate results. When all intermediate
results are re-computed at each output
pixel (sch1), it requires
only a small amount of intermediate buffering
(3 memory tiles), but a huge amount of
compute logic (769 PEs). The schedule
sch2 shows a middle point where some intermediate
results are buffered, and in sch3 all
intermediate results are buffered to avoid any
recomputation. These results show that it is beneficial to buffer the intermediates for Harris to drastically reduce the PE requirements with a few more memories.

Furthermore, we can unroll compute kernels to increase the throughput. As shown for sch4 of~\autoref{tab:sch-exploration}, unrolling doubles throughput and roughly doubles resource utilization.

In sch5 of~\autoref{tab:sch-exploration} we use \texttt{tile} with parameters that are 2 times larger in each dimension. The increased tile size lets the accelerator run for more cycles and compute a larger portion of the output. In this case, the number of memories did not change, since our MEM tiles were under-utilized previously. 

We can use \texttt{hw\_accelerate} to specify what part of the application is performed on the accelerator; the rest executes on the host processor. In our last schedule, sch6, we move the last stage of Harris to the host processor. This reduces the PE and MEM requirements of our application. Using Halide's succinct scheduling language and our compiler system, we are able to explore these six designs with little design effort.

\begin{table}[t]
    \caption{
        Compiler results for Harris application with different Halide schedules. A high-level scheduling decision explores different resource and throughput requirements.
        \label{tab:sch-exploration}
    }
    \centering
    \begin{tabular}{p{0.30\colw} p{0.15\colw} p{0.09\colw} p{0.12\colw} p{0.10\colw}} \toprule
         Harris Schedule      &  Output \newline Pixels/Cycle & \# PEs  & \# MEMs  & Runtime (cycles)   \\
         \midrule
         sch1: recompute all         & 1   & 769     & 3      & 4097    \\
         sch2: recompute some & 1   & 145     & 5    & 4103     \\
         sch3: no recompute      & 1   & 83      & 5    & 4146   \\
         sch4: unroll by 2     & 2   & 194     & 10      & 2154    \\
         sch5: 4x larger tile  & 1   & 85      & 5     & 16434     \\
         sch6: last stage on CPU    & 1   & 67      & 4      & 4142     \\
         \bottomrule
    \end{tabular}
\end{table}

\subsection{Effectiveness of Memory Mapping Optimizations}

Our memory mapping algorithm is designed
to maximize compute utilization while
also keeping application latency and
memory use low. ~\autoref{tab:schedule_speedup} shows
the effect of our hardware pipeline
scheduling algorithm on application
completion time compared to a na\"ive
strategy that executes each kernel
sequentially and does not pipeline
any of the loops. Speedups range from
around $3\times$ for resnet to $22\times$ for harris
and camera. Gaussian, harris,
unsharp, and camera all
have very different numbers of stages,
which is reflected in the
fact that in sequential scheduling
harris takes more than three times
as many cycles as gaussian. Using our
scheduling algorithm, all four
of these applications have completion
times that are nearly identical, because
with appropriate pipelining the
latency cost of an additional stage
in a stencil pipeline is proportional
to the number of additional lines of the image
that the stage must wait for to start,
while without cross-stage pipelining
the cost of the stage is proportional
to the latency of the entire stage.
For applications that do not lend
themselves to fine-grained cross-stage pipelining
(such as resnet) the speedup is
less dramatic, but still noticeable
at $3\times$.

\autoref{tab:memory_reduction} shows
the effect of our scheduling optimizations
on locality by measuring the required
SRAM capacity (in 16 bit words) for the
sequentially scheduled applications
versus the applications scheduled
using our pipeline scheduler.
Here the results are similar to the
results for overall completion time.
The image processing pipelines see dramatic
decreases in memory requirements of
between $28\times$ and $300\times$. This is
because the working sets of stencil
operations when appropriately pipelined
are only a few lines of the input image
for each stage. But when stages
are executed sequentially, the inter-stage
buffers must be large enough to hold
the entire output image created by the
producer stage. Mobilenet sees a similar,
but less dramatic decrease in its
memory requirements since it is
structurally similar to a stencil
pipeline. Resnet sees no change in
its memory demands since adjacent stages cannot be fused, so it only pipelines
the individual kernels and then
executes each one sequentially.

\begin{table}[t]
    \centering
    \caption{
        The effect of our scheduling optimizations compared to a na\"ive strategy that sequentially schedules all loop nests.
        \label{tab:schedule_speedup}
    }
    \begin{tabular}{p{0.18\colw} p{0.2\colw} p{0.26\colw} p{0.10\colw} } \toprule
        Application & 	Sequential Completion Time (cycles)&	Optimized Schedule Completion Time (cycles) & Speedup\\
         \midrule
         gaussian &	27159 &	4102 & 6.62      \\
         harris &	92227 &	4120 & 22.39     \\
         upsample	& 53247 &	16387 & 3.25 \\
         unsharp &	49279 &	4119 & 11.96     \\
         camera &	92013 &	4122 & 22.32     \\
         resnet &	44876 &	15614 & 2.87     \\
         mobilenet &	22463 &	1026 & 21.89 \\
         \bottomrule
    \end{tabular}
\end{table}

\begin{table}[t]
    \caption{
        The effect of our hardware pipeline scheduling optimizations on required SRAM capacity.
        \label{tab:memory_reduction}
    }
    \centering
    \begin{tabular}{p{0.15\colw} p{0.20\colw} p{0.15\colw} p{0.15\colw} } \toprule
        Application & Sequential Schedule SRAM Words &	Final SRAM Words & Memory Reduction Factor \\
         \midrule
        gaussian &	11784 &	128 & 92.06  \\
        harris &	41080 &	640 & 64.19  \\
        upsample &	20480 &	67  & 305.67 \\
        unsharp &	23584 &	834 & 28.28  \\
        camera &	37972 &	518 & 73.31  \\
        resnet &	14048 &	14048 & 1.00  \\
        mobilenet &	9136 &	1240 & 7.37  \\
         \bottomrule
    \end{tabular}
\end{table}

\section{Related Work}

Creating tools to support application-tailored accelerators is an active area of research, and our system builds on much of this work.  
We break this work into three areas: compiler frameworks, push memory abstractions, and domain-specific accelerator generators.

\textbf{Compiler Frameworks}
Neither conventional software compilers nor existing
hardware compilers are well-suited to target
push memories.
Conventional compilers for imperative programming languages that assume a von Neumann machine,
such as LLVM, are built around an intermediate
representation that separates control, data
access, and arithmetic. That is, they assume that the
most important piece of
architectural state that the compiler
must manage is the register file. When using efficient push memories, however, memory, address generation, and control are
grouped into a single unit that the compiler must generate \textit{instructions} for, and then connect its ports to the compute units.

High level synthesis (HLS) tools
such as Vivado~\cite{vivado_manual}, LegUp~\cite{canis2011legup}, Catapult~\cite{catapult_manual} and others~\cite{maxcompiler, altera_opencl}, are designed
to solve scheduling and resource binding problems at a finer granularity than those seen when compiling to push memories.
HLS tools compile C or C++
programs using conventional software compiler
frontends and intermediate representations (IRs)
(the Edison design group C++ frontend
for Catapult and clang+LLVM for Vivado and
LegUp). They then schedule the instructions in
the standard software IR, assign instructions
to functional units, and emit code.
This strategy works well when
targeting FPGAs or ASIC technology
libraries, because the
architectural primitives (such as registers and
LUTs) are more fine-grained than the
instructions in the compiler IR that is scheduled.
When compiling to programmable push memory accelerators, where
the architectural primitives are much more
coarse-grained than a typical RISC instruction,
this approach does not work.
Academic languages
such as Spatial \cite{koeplinger2018spatial} and HeteroCL \cite{lai2019heterocl} provide a more abstract programming
model for accelerators, but offload the work
of defining the memory micro-architecture to the
user. Though HeteroCL uses a unified DSL frontend to describe their memory optimization and spatial architecture, their backend implementation still depends on separate templates. Whereas, our compilation is general enough to map both stencil applications and deep neural networks onto hardware accelerator with unified buffers. 

While HLS tools can translate the Halide IR directly to hardware, they do not support the memory optimizations we describe. 
Modern HLS tools such as Vivado HLS or Catapult are well
suited to arithmetic mapping and
exploiting pipeline parallelism within
the bodies of individual loops \cite{meeus2012overview}.
However, they perform limited memory \cite{pu2017programming}
and cross-loop optimizations \cite{zuo2013improving}. As a result,
they are not good at exploiting pipeline parallelism
across different loop nests in a computation,
and require a great deal of manual effort by
users to create high quality code for deep pipelines \cite{lai2019heterocl}.

The success of HLS tools has led
to a generation of high-level compilers for hardware. These include
polyhedral compilers which perform source-to-source
transformations on HLS C code \cite{pu2017programming,zuo2013improving}
and others which start from high level
DSLs and emit HLS C++ such as Halide-HLS
\cite{pu2017programming}, Hetero-Halide
\cite{li2020heterohalide}, and
PolyMage-FPGA \cite{chugh2016dsl}. While these
are important contributions to FPGA and ASIC
toolchains, they depend on a conventional HLS
tool as a code generator, and thus do not address
the problem of targeting a programmable 
push memory accelerator. Our approach adds
a unified buffer abstraction, which enables
our compiler to use polyhedral analysis to
map to accelerators using push memories.

\textbf{Push Memory Abstractions} Our unified buffer borrows from Buffets~\cite{buffets}, a buffer implementation idiom that can be reused across multiple domains with explicit data orchestration. Buffets are a hardware primitive, not a compiler abstraction, while our unified buffer is both.
Our hardware includes addressing and sequencing control, which enables a compiler to better optimize the storage requirements before being mapped to hardware. It also enables a set of optimizations when generating the hardware. In addition, we reduce manual design effort and improve productivity by using systematic analysis conducted by our tool chain to extract and optimize the buffer parameters used in the unified buffer. 
Plasticine \cite{plasticine} also supports push memories by creating addressing units associated with their memories. Spatial \cite{koeplinger2018spatial} provides a high-level programming language for this push-memory architecture, but requires users to explicitly orchestrate data movement between different memories.
Nowatzki et al. proposed a low-level programming model for stream dataflow accelerators \cite{nowatzki2017stream}. Since their memory architecture is a global scratchpad, which is not distributed across the chip, their memory ISA contains dynamic scheduling which may not suitable for push memory accelerators. Moreover, our access pattern is more general, which can support an arbitrary number of dimensions. 

\textbf{Domain-Specific Accelerator Generators} There has also been work to automate domain-specific accelerator design. 
Image processing accelerator generation languages such as Darkroom \cite{Hegarty:2014:DCH:2601097.2601174}, Rigel \cite{hegarty2016rigel}, Aetherling \cite{durst2020type}, Hetero-Halide \cite{li2020heterohalide}, HIPACC-FPGA \cite{reiche2014code}, PolyMage-FPGA \cite{chugh2016dsl}, SODA \cite{chi2018soda} and Halide-HLS \cite{pu2017programming} can automatically generate FPGA implementations of image processing pipelines. These systems either target FPGAs, or ASICs, which either have large overheads, or are inflexible. 

To efficiently execute DNNs, Zhang et al.~\cite{zhang2015optimizing} optimize DNN data blocking using double buffer structures and synthesize a pipelined FPGA accelerator from Caffe~\cite{jia2014caffe}. DNNWeaver~\cite{sharma2016high} also generates synthesizable designs automatically from Caffe, with support for more types of layer implementations. DNNBuilder~\cite{Zhang:2018:DAT:3240765.3240801} proposes a fine-grained layer-based pipeline architecture with a line-buffer-based scheme to reduce FPGA on-chip memory usage. However, such library-based frameworks heavily rely on their backend implementation. With the architectures being pre-determined, extending them to support fast-moving domains would require significant development effort from domain experts. VTA~\cite{Moreau:2018:LVH:3229762.3229766, moreau2018vta} provides a full hardware-software stack for DNN acceleration using a modified version of Halide IR. It proposes an ISA to map DNN layers onto optimized operators and offloads these computations to their proposed FPGA accelerator. VTA uses a CPU-style pull memory with instruction fetching, which adds both computation and energy overhead. In contrast, our unified buffer accelerator generator offloads memory analysis to the software compiler to generate an optimized hardware memory design and computation pipeline.
These domain-specific hardware generators reduce design effort when building ASIC or FPGA designs even more than classical HLS, but like classical HLS tools they cannot target programmable accelerators based on push memories.

\section{Conclusion}

In this paper, we have presented a new abstraction for push memories which supports efficient hardware realization. This abstraction---unified buffer---presents a tractable target for an optimizing compiler, and we describe one such compiler.
We believe that our work will become increasingly important as push memory accelerators are required to improve power and performance in image processing, machine learning, and other dense linear algebra programs.

\bibliographystyle{plain}
\bibliography{references}

\end{document}